\documentclass[12pt,a4paper]{book}
\usepackage[left=1.25in, right=1in, top=1in, bottom=1in]{geometry}

\usepackage{graphicx}
\usepackage{url}
\usepackage{emptypage}
\graphicspath{ {Images/} }
\usepackage{float}
\usepackage[colorlinks=true,linkcolor=blue]{hyperref}
\restylefloat{table}
\usepackage{amsfonts, amsmath, amssymb}
\usepackage{subcaption}
\usepackage{amsmath}
\usepackage[table,xcdraw]{xcolor}
\usepackage{imakeidx}
\usepackage{afterpage}
\usepackage{pdfpages}
\newcommand\blankpage{%
    \null
    \thispagestyle{empty}%
    \newpage}
    
\makeindex

\begin{document}
\begin{titlepage}
   \begin{center}
       \vspace*{1.5cm}
    
       \textbf{\Huge Simulations of Bosonic BMN Matrix Model}
 
      \large
       \vspace{3.5cm}
 
       \large{\textbf{Adeeb Mev}}
 
       \vspace{2.5cm}
 
       \textit{\large A dissertation submitted for the partial fulfillment of\\
       BS-MS dual degree in Science}
 
       \vspace{3.8cm}
 
       \includegraphics[scale=0.5]{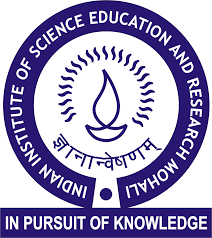}
        
       \textbf{Department of Physical Sciences \\ Indian Institute of Science Education and Research Mohali\\
       June 2020}

   \end{center}
\end{titlepage}
\afterpage{\blankpage}
\clearpage
\thispagestyle{empty}
\newgeometry{left=1.25in, right=1in, top=1in, bottom=1in}
\clearpage
\thispagestyle{empty}
\includepdf{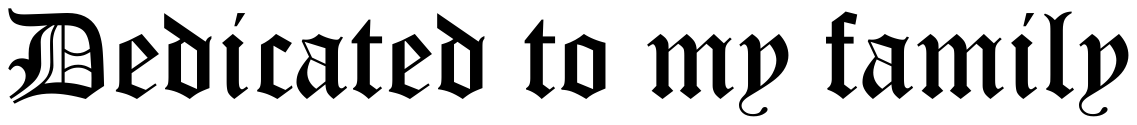}
\afterpage{\blankpage}
\clearpage
\afterpage{\blankpage}
\parbox{5.5in}{
\begin{center}
\Large \textbf{Certificate of Examination}
\end{center}
\small{ 
\par

\noindent This is to certify that the dissertation titled 
\textbf{Simulations of Bosonic BMN Matrix Model}
submitted by \textbf{Adeeb Mev} (Reg. No. MS15058) for the partial fulfillment of BS-MS dual degree 
programme of the Institute, has been examined by the thesis committee duly appointed by the
Institute. The committee finds the work done by the candidate satisfactory and recommends that the
report be accepted.
}
\vspace{2.0cm}
\begin{flushright}
 Dr. Manabendra Nath Bera  \hspace{0.5cm} Dr. Ambresh Shivaji \hspace{0.5cm} Dr. Anosh Joseph
 
 (Supervisor)
\end{flushright}
\vspace{1.5cm}
\begin{flushright}
Dated: June 14, 2020 
\end{flushright}
}

\thispagestyle{empty}
\clearpage
\thispagestyle{empty}
\parbox{5in}{
\begin{center}
\Large \textbf{Declaration}
\end{center}
\small{ 
\par\noindent The work presented in this dissertation has been carried out by me under the guidance of 
Dr. Anosh Joseph at the Indian Institute of Science Education and Research (IISER) Mohali.\newline

\noindent This work has not been submitted in part or in full for a degree, a diploma,
or a fellowship to any other university or institute. Whenever contributions
of others are involved, every effort is made to indicate this clearly, with due
acknowledgement of collaborative research and discussions. This thesis is a
bonafide record of original work done by me and all sources listed within
have been detailed in the bibliography.
}
\vspace{1cm}
\begin{flushright}
 Adeeb Mev
 
 (Candidate)
\end{flushright}

\begin{flushright}
 Dated: June 14, 2020
\end{flushright}
\small{ \par\noindent In my capacity as the supervisor of the candidate's project work, I certify
that the above statements by the candidate are true to the best of my
knowledge.}

\vspace{1.5cm}
\begin{flushright}
 Dr. Anosh Joseph
 
 (Supervisor)
\end{flushright}
}

\clearpage
\afterpage{\blankpage}
\clearpage
\thispagestyle{empty}
\parbox{5.5in}{
\begin{center}
\Large \textbf{Acknowledgments}
\end{center}
\small{\par\noindent I express my sincere gratitude to my supervisor, Dr. Anosh Joseph for his patience, motivation, prudent comments, valuable suggestions, and beneficial information, which have helped me remarkably in my research and in writing this thesis. His enormous knowledge and thorough experience in lattice field theory have enabled me to complete this research successfully. I am incredibly thankful to him for sparing his precious time to guide me, clarifying my queries, correcting me during my research and sharing his workstation without which simulations would have never been completed.\newline

\noindent Besides my advisor, I wish to thank the rest of my thesis committee members, Dr. Manabendra Nath Bera and Dr. Ambresh Shivaji, for their insightful comments and encouragement.\newline

\noindent I also extend my sincere thanks to my friend Nikhil Tanwar, who has helped me in improving my research and programming skills.\newline

\noindent I acknowledge IISER Mohali for providing me with the best infrastructure and environment for carrying out this project. I am also thankful to the Department of Science and Technology (DST), Government of India, for supporting me with the institute fellowship during the past five years.\newline

\noindent To conclude, above all, it was the Almighty and my family who have been there all these years actively supporting and guiding me in life.\newline
}}
\clearpage
\afterpage{\blankpage}
\clearpage
\pagenumbering{roman}
\listoffigures \addcontentsline{toc}{chapter}{List of Figures}
\clearpage
\listoftables \addcontentsline{toc}{chapter}{List of Tables}
\cleardoublepage
\clearpage
\tableofcontents
\clearpage
\begin{center}
 \Large \bf Abstract
\end{center}

\noindent In this thesis we provide the results obtained through lattice Monte Carlo simulations of the bosonic BMN and the bosonic BFSS matrix models. The simulations are performed using Hybrid Monte Carlo (HMC) algorithm. The BMN matrix model is expected to have a Hagedorn/deconfinement type phase transition as the temperature is varied in the system. The Polyakov loop is used as an order parameter for detecting the phase transition. Besides the Polyakov loop, other prime observables such as the internal energy and the extent of space were also computed. We also check the validity of numerical simulation algorithms by exploring the behavior of various relevant toy models. \\

\noindent As the main result of this thesis, we present a parametrized phase diagram of the bosonic BMN matrix model constructed using two dimensionless parameters: a dimensionless coupling constant and a dimensionless temperature. \addcontentsline{toc}{chapter}{Abstract}
\clearpage
\mainmatter
\chapter{Introduction}
\section{String theory and M theory}

String theory is a field in theoretical and mathematical physics, which represents a major dream of theoretical physicists towards building `a theory of everything.' It is a set of attempts made to unify gravity with other three fundamental forces in Nature: electromagnetic, strong nuclear and weak nuclear forces. The basic idea of this theory is that all matter particles and force mediators are made up of one-dimensional objects called strings. The theory explains how these strings propagate in spacetime and interact with each other. It replaces the point like particles, described as the fundamental entities in the Standard Model, by these one-dimensional objects. In string theory, each elementary particle corresponds to a unique vibrational pattern of a string and these vibrational patterns determine their properties like charge, mass, spin, etc. \\

There are five types of string theories: type I, type IIA, type IIB and two flavors of heterotic string theory (SO(32) and $E_8\times E_8$). These theories differ depending on whether strings are open or closed, or they are oriented or unoriented. Each of these theories lives in ten spacetime dimensions. \\

In 1995, Edward Witten came up with a new theory in which he suggested that all five string theories are part of a vast, yet still undiscovered 11-dimensional theory known as M theory. The M is undefined and may stand to refer to a ``membrane'' or ``matrix.'' \\

String theory was first studied in the late 1960s and since then it has been evolving. The theory has enriched enormously the fields of quantum gravity, high energy physics, nuclear physics, condensed matter physics, and pure mathematics. However, there still exists a debate over its validity since no part of this theory has been verified experimentally. \\

\section{The BFSS matrix model}

The BFSS matrix model is a one-dimensional supersymmetric Yang-Mills theory. It was conjectured in 1996 by T. Banks, W. Fischler, S.H. Shenker, and L. Susskind \cite{Banks_1997}. This model is regarded as a low energy effective description of $N$ $D0$-branes of type IIA superstring theory \cite{Filev_2016}. It is speculated that in the large-$N$ limit this model is related to the uncompactified eleven-dimensional M theory. One way to obtain this model is to dimensionally reduce the ten-dimensional supersymmetric Yang-Mills theory to one-dimension. The resultant reduced action is 
\begin{equation}
S = \frac{1}{2 g_{\rm YM}^2} \int d t \, \operatorname{Tr}\left[\left(D_{t} X^{i}\right)^{2} + \psi^{\alpha} D_{t} \psi^{\alpha}+\frac{1}{2}\left[X^{i}, X^{j}\right]^{2}+i \psi^{\alpha} \gamma_{\alpha \beta}^{j}\left[\psi^{\beta}, X^{j}\right]\right],
\end{equation}
where $g_{\rm YM}$ is the Yang-Mills coupling, $D_t = \partial_t - i [A, \cdot]$ represents the covariant derivative, the indices $i , j = 1, \cdots, 9$ run over the scalars of the theory, and spinor indices $\alpha, \beta = 1, \cdots,16$. All degrees of freedom are $N\times N$ Hermitian matrices.

\section{The BMN matrix model}

The BFSS matrix model is defined on a flat Minkowski spacetime, but matrix models can also be considered on curved spacetime. 

The maximally supersymmetric pp-wave background, which preserves 32 supercharges, gives an example of this type. Berenstein, Maldacena and Nastase \cite{Berenstein_2002} proposed the BMN matrix model on this background \cite{Kawahara_2006}. This model is constructed by deforming the BFSS matrix model through the addition of a mass parameter.

This model has fuzzy spheres as classical solutions due to presence of mass and Myers terms \cite{Myers_1999}. The presence of mass term leads BFSS's $SO(9)$ global symmetry to break down to $SO(6)\times SO(3)$. The action of PWMM (plane wave matrix model) or BMN has the form  
\begin{eqnarray}
S &=& \frac{1}{2 g_{\rm YM}^{2}} \int d t \, \operatorname{Tr} \left[ \left(D_{t} X^{i}\right)^{2}+\psi^{\alpha} D_{t} \psi^{\alpha}+\frac{1}{2}\left[X^{i}, X^{j}\right]^{2}+i \psi^{\alpha} \gamma_{\alpha \beta}^{j}\left[\psi^{\beta}, X^{j}\right] \right. \nonumber \\
&& \left. -\frac{\mu^{2}}{3^{2}}\left(X^{I}\right)^{2}-\frac{\mu^{2}}{6^{2}}\left(X^{I^{\prime}}\right)^{2}-\frac{\mu}{4} \psi^{\alpha}\left(\gamma^{123}\right)_{\alpha \beta} \psi^{\beta}-i \frac{2 \mu}{3} \epsilon_{I J K} X^{I} X^{J} X^{K} \right],
\end{eqnarray}
where $\mu$ is the deformation parameter, and 
\begin{equation*}
\begin{aligned}
i , j = 1, \cdots, 9, \\ 
\alpha, \beta = 1, \cdots, 16, \\ 
I , J , K = 1, 2, 3, \\ 
I' = 4, \cdots,9, \\ 
\gamma^{123} = \frac{1}{6} \epsilon_{I J K} \gamma^{I} \gamma^{J} \gamma^{K}, \\
D_t = \partial_t - i[A, \cdot].
\end{aligned}
\end{equation*}

\section{Advantages of BMN model over BFSS model}

There are certain advantages of the BMN model over the BFSS model. Firstly, the BMN model has a discrete energy spectrum and a well defined canonical ensemble whereas the canonical ensemble of the BFSS model does not exist \cite{Costa_2015} due to the presence of the so-called flat directions. The existence of flat directions means that the eigenvalues of commutating matrices $X^i$ can attain arbitrarily large values without costing energy. Flat directions in the BFSS model leads to divergences and non-existence of the partition function defined at a finite temperature. However, Monte Carlo simulations of the BFSS model can be accomplished thanks to the existence of a meta-stable thermal equilibrium, which has a decay rate that is very small at large $N$. Flat directions are absent in the BMN model because of mass terms. \\
 
Secondly, the BMN model has two dimensionless parameters - a dimensionless coupling constant 
\begin{equation}
g \equiv \frac{\lambda}{\mu^3},
\end{equation}
with $\mu$ denoting a mass parameter and $\lambda$ the `t Hooft coupling; and a dimensionless temperature
\begin{equation}
	t = \frac{T}{\mu},
\end{equation}
with $T$ denoting the (dimensionful) temperature.

These two parameters can be used to parametrize a two-dimensional phase diagram of the model. This signifies that the dual gravitational description at large $N$ and strong coupling $g \gg 1$ can be used to predict various observables as functions of $t$ \cite{Joseph_2015}. \\

Thirdly, the model becomes weakly coupled in the limit $\mu \to \infty$, so it can be studied perturbatively \cite{Dasgupta_2002}.

\section{BMN matrix model and black holes}

One of the principal reasons to investigate the BMN model is the connection of matrix models with black hole states. The BFSS model, at finite temperature, is related to the black hole state of type IIA supergravity \cite{R_Klebanov_1998} \cite{Banks_1998}. The connection between BFSS matrix model and black hole states is also established by Kabat and Lowe in Ref. \cite{Kabat_2001}. They calculated the entropy of the quantum mechanical system that agrees well with the Beckenstein-Hawking entropy of a ten-dimensional non-extremal black hole. The free energy of the black hole in terms of the parameters of the gauge theory (the BFSS model) can be written as \cite{SEMENOFF_2004}
\begin{equation}
\label{eq:internal-energy-formula}
\frac{F}{T} = -4.115 N^{2} \left(\frac{T^{3}}{g_{\rm YM}^{2} N}\right)^{3 / 5}.
\end{equation}

The formula given in Eq. \eqref{eq:internal-energy-formula} is of interest for a couple of reasons. Firstly, the dependence of the free energy on the 't Hooft coupling ($\lambda=g^2_{\rm YM} N$). The 't Hooft large-$N$ limit corresponds to the region where supergravity description is valid and therefore, is appropriate to look for the behavior of the black hole in the BFSS model. Secondly, the $N^2$ dependence of the free energy, which is also seen in the deconfined state of a gauge theory system. \\

It would be significant to reproduce the above formula using calculations in the matrix model. To get the behavior mentioned above in the matrix model, one must look in the strong coupling limit. But in such limits, perturbation theory cannot be applied. Thus any analytic derivation for this expression has not been made. The other approach is to use techniques based on numerical simulations. It is a decent alternative as it does not have any such restrictions.

Further, from the study of black holes dual to the deconfined phase of the BMN model, the critical temperature in the strong coupling limit was determined in Ref. \cite{Costa_2015} 
\begin{equation}
\lim_{g \to \infty} \frac{T_{c}(g)}{\mu} = 0.105905(57),
\end{equation}
which is again very difficult to derive analytically in the matrix model. \\

The authors of Ref. \cite{Costa_2015} also predicted the parametrized phase diagram for the BMN model, and it is shown in Fig. \ref{fig:1}. In the opposite limit of weak coupling, that is, in the $g \to 0$ limit, the critical temperature was predicted using perturbative calculations in Refs. \cite{furuuchi2003fivebrane} and \cite{Hadizadeh_2005}
\begin{eqnarray}
\left.\lim_{g \to 0} \frac{T}{\mu}\right|_{\mathrm{c}} &=& \lim_{g \to 0}\frac{1}{12 \log 3} \left[ 1 + \frac{2^{6} \cdot 5}{3} g  \right. \nonumber \\
&&~~~~~~ \left. - \left(\frac{23 \cdot 19927}{2^{2} \cdot 3} + \frac{1765769}{2^{4} \cdot 3^{2}} \log 3\right) g^{2} + \mathcal{O} \left(g^{3} \right) \right] \nonumber \\
&\approx&  0.076.
\end{eqnarray}

It would be a significant achievement if we could verify the above results using numerical non-perturbative methods. We will resort to Monte Carlo simulations to study the BMN model numerically.

\section{The deconfinement phase transition}

For systems owning an exponentially increasing density of states, 
\begin{equation} 
\begin{aligned} \rho(E) \sim e^{\beta_H E},    
\end{aligned}\
\end{equation} 
there exists an upper limiting temperature above which the partition diverges and no longer exists,
\begin{equation}
    \begin{aligned}
     \lim_{T \to T^-_H} \operatorname{Tr} \left[ e^{-\beta H} \right] \to \infty,
    \end{aligned}
\end{equation}
such a limiting temperature is known as the Hagedorn temperature \cite{SEMENOFF_2004}. \\

Above the cutoff temperature $T_H$, the partition function does not exist. However, its existence can be made by holding $N$ large but finite. Doing so would break off the exponential growth in the asymptotic density of states at some large energy value. At temperatures higher than $T_H$, the entropy and the energy are governed by the states at and above the cutoff scale, the free energy jumps from $\mathcal{O}(1)$ to $\mathcal{O}(N^2)$. Thus \cite{Hadizadeh_2005}
\begin{equation}
\begin{array}{l}
\lim _{N \to \infty} \frac{F}{N^{2}} = 0 \quad \text {(confined)}, \\
\lim _{N \to \infty} \frac{F}{N^{2}} \neq 0 \quad \text {(deconfined)}.
\end{array}
\end{equation}

The transition from confined to deconfined state of the gauge theory is called the deconfinment phase transition. In the confined phase, state before the deconfinement phase transition, the quantum states of the Hamiltonian must be singlets under the gauge symmetry \cite{SEMENOFF_2004}. This condition breaks as soon as the system reaches the Hagedorn temperature.\\

The deconfinement phase transition has been found in large $N$ gauge theories such as weakly coupled Yang-Mills theory \cite{Aharony_2004} \cite{Sundborg_2000}. It is expected that this type of phase transition exists in the BFSS and the BMN matrix models also \cite{SEMENOFF_2004} \cite{Hadizadeh_2005}. \\

\begin{figure}[ht!]
\begin{center}
\includegraphics[scale=0.5]{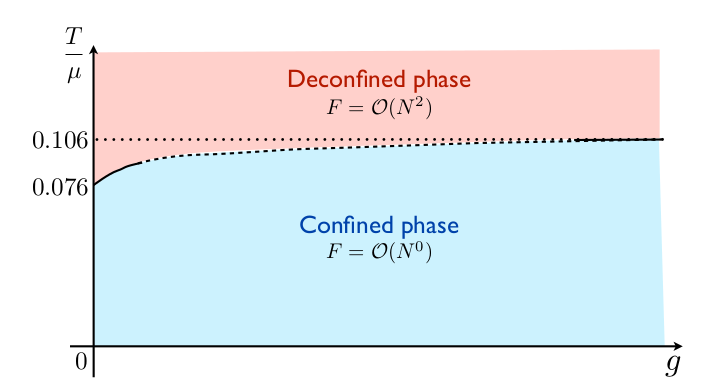}
\end{center}
\caption[The phase diagram of the BMN matrix model.]{The phase diagram of the BMN matrix model. At low temperature the system is in a confined phase where the free energy scales as $N^0$. As the temperature increases, the system undergoes a first-order phase transition to a deconfined phase where the free energy scales like $N^2$. (This figure is taken from Ref. \cite{Costa_2015}.)}
\label{fig:1}
\end{figure}

The deconfinement transition in the matrix models is associated with the spontaneous breaking of the centre symmetry, $A(t) \mapsto A(t) + c \mathbf{1}$. There is an order parameter, called the Polyakov loop, for this type of symmetry breaking \cite{POLYAKOV1978477}. It is defined as the trace of the holonomy of the gauge field around the finite temperature Euclidean time circle. 
\begin{equation}
P = \frac{1}{N} \operatorname{Tr} \mathcal{P} \left[\text{exp} \left(i \oint A\right) \right].
\end{equation}
This operator is gauge invariant. Its expectation value, which is zero in the confined phase, jumps to a non-zero value in deconfined phase. This is because the Polyakov loop is a unitary matrix and its eigenvalues in the confined phase are uniformly distributed on a unit circle, whereas, in the deconfined phase, they start to clump together.
\begin{equation}
\begin{aligned}
&\langle P \rangle = 0 \quad \text {(confined phase)}, \\
&\langle P\rangle \neq 0 \quad \text {(deconfined phase)}.
\end{aligned}
\end{equation}
It is widely used to study the deconfinement type phase transitions in higher dimensional gauge theories. We also use this observable as an order parameter in the simulations carried out in this work.

The phase diagram shown in Fig \ref{fig:1} has been recently recovered using Monte Carlo simulations in Ref. \cite{schaich2020thermal}. They used the Polyakov loop as an order parameter to trace the transition temperature. \\

The chapters of this thesis are arranged in a series of steps towards building up the numerical simulation algorithms used to simulate the BFSS and BMN matrix models. Each chapter includes a specific topic. In Chapter 2, we discuss the discretization of the matrix models on a Euclidean lattice. The basics of Monte Carlo integration along with two algorithms, Metropolis and Hamiltonian Monte Carlo (HMC), are discussed in Chapter 3. In Chapter 4, we present the Monte Carlo simulations of a toy model, the harmonic oscillator with potential $\mu^2 X^2$. It is an exactly solvable model, and thus serves as a good starting point to test out the numerical integration methods. After this, the numerical simulations of the model containing a commutator potential term $[X^i, X^j]^2$, is covered in Chapter 5. In Chapter 6, a gauge field is included in the model, and integration over gauge variables is investigated using the $D = 4$ model. The $D = 4$ model is a toy model of the BFSS matrix model, and therefore is suitable for examining the behavior of the BFSS matrix model. Finally, the bosonic BFSS and the bosonic BMN matrix models are investigated in Chapters 7 and 8, respectively. In Chapter 9 we provide conclusions.

\chapter{Lattice Discretization}
\section{The quenched model}

The primary focus of this thesis is to perform numerical simulations of the bosonic BMN matrix model. Therefore, the first step is to remove the fermions from the model. Inclusion of fermions leads to various challenging computational difficulties. Firstly, there exists the problem of fermion doubling: placing fermions on a Euclidean spacetime lattice leads to the fermions to behave like multiple degenerate particles. A solution to this difficulty requires the addition of a term, known as the Wilson term, directly into the action of the theory. This term decouples the degenerate states but comes with further challenges. Secondly, the fermionic terms in the action can be integrated out. However, doing so would leave a complex fermion determinant, which would pose problems when we use importance sampling based Monte Carlo algorithms. The fermion determinant has a rapidly fluctuating phase which weakens the probabilistic interpretation of the Euclidean action in the path integral. It is also in general time consuming to evaluate the determinant at every simulation time step during the molecular evaluation. Therefore, in general, the inclusion of fermions in the lattice field theory simulations is a difficult task. \\

Also, at high temperatures, fermions decouple from the full action and leave only the remnant bosonic behavior. Thus, at high temperatures, the system should be described by just the bosonic action. \\  

Therefore, dropping fermions is a step taken forward to make the system simpler to examine. Removing fermions completely from an action is also known as fermion quenching.\\ 

Simulating the bosonic part of the action, the quenched model, is a complicated problem in itself, and that will be the focus of this work. We have strong reasons to believe that in the absence of fermions, the bosonic part of the model, at high temperature, would show the same behavior as that of the full model.

\section{Euclidean action and partition function}

The thermodynamics of the matrix model can be investigated using the Euclidean action compactified on a temporal circle. An action in the Minkowski space is moved into the Euclidean space through Wick rotation. The process of Wick rotation involves the substitution of the Minkowski time for the imaginary Euclidean time ($t = - i t_E$). This just alters the sign of a few terms in the action. The system can be studied at a finite temperature by compactifying the action on a circle with the circumference $\beta \equiv 1/T$. The compactified Euclidean bosonic actions for both the models are

\begin{equation}
\begin{aligned}
S_{\mathrm{E}}^{\rm BFSS} = \frac{N}{\lambda} \int_{0}^{\beta} d t \operatorname{Tr}\left[\frac{1}{2}\left(D_{t} X^{i}\right)^{2}-\frac{1}{4}\left[X^{i}, X^{j}\right]^{2}\right],
\end{aligned}
\end{equation}

\begin{eqnarray}
S_{\mathrm{E}}^{\rm BMN} &=& S_{\mathrm{E}}^{\rm BFSS} + \frac{N}{\lambda} \int_{0}^{\beta} d t \operatorname{Tr} \left[ \frac{1}{2}\left(\frac{\mu}{3}\right)^{2}\left(X^{I}\right)^{2} \right. \nonumber \\
&& ~~~~~~~~ \left. + \frac{1}{2}\left(\frac{\mu}{6}\right)^{2}\left(X^{I'}\right)^{2} + i \frac{\mu}{3} \epsilon_{I J K} X^{I} X^{J} X^{K} \right],
\end{eqnarray}
where $D_t = \partial_t - i[ A, \cdot]$ is the covariant derivative and the boundary conditions are periodic in time, $X^i(t) = X^{i}(t + \beta)$ and $A(t) = A(t + \beta)$. The coupling $g_{\rm YM}^2$ is written in the form of `t Hooft coupling to make the behavior of the system invariant for distinct values of $N$.\\

Using the Euclidean action, the partition function is written as
\begin{equation}
Z = \int \mathcal{D}[A] \mathcal{D}[X] e^{-S_{E}[X]}.
\label{eq:1}
\end{equation}
It counts all accessible states of a system and is used to define fundamental quantities in thermodynamics like free energy, entropy, etc. \\

The free energy has the expression
\begin{equation}
\mathcal{F} \equiv-\frac{1}{\beta} \ln Z.
\end{equation}
Unfortunately, the free energy cannot be obtained easily using Monte Carlo simulations as it demands the evaluation of the partition function explicitly. However, we can make use of another useful quantity, defined by
\begin{equation}
E \equiv \frac{d}{d \beta}(\beta \mathcal{F}) = -\frac{d}{d \beta} \log Z.
\end{equation}

It will be interesting to look at the behavior of this quantity in the matrix model simulations.

\subsection{Connection to statistical mechanics}

There is a structural equivalence between the Euclidean path integral of a lattice field theory and the partition function of a statistical mechanical system. 

To understand this, let us consider a spin system in statistical mechanics. Assuming that the spins are distributed on a lattice, the partition function is written as 
\begin{equation}
Z_s = \sum_{\{s\}} \mathrm{e}^{-\beta H[s]},
\end{equation}
where the sum is over all possible spin configurations. \\

Equation \eqref{eq:1} has a similar structure to that of the above equation. The weight factor $e^{-\beta H}$ is replaced by $e^{-S_{E}}$ and the sum over the spin configurations are replaced by a path integral. This structural equivalence suggests that performing simulations of a quantum field theory system by implementing it on a spacetime lattice, is the statistical mechanical study of the system. It also provides an advantage to lattice field theory: it can make use of the analytical and numerical methods developed within statistical mechanics. These methods are discussed in the next chapter.

\section{Gauge fixing}

The lattice simulation of the model is possible even without fixing the gauge, but it would be computationally expensive. Actually, the gauge symmetry in the model results in redundant dynamical variables and including these variables in the lattice action would make the simulation inefficient and time consuming. Therefore, it is essential to remove them before embarking on simulations. \\

\section{The link variables}

In general, a continuum action that is invariant under a set of gauge transformations may not remain invariant when it is discretized on a lattice. This occurs when the action has a derivative term, which on discretization, leaves two field variables at different lattice sites, thus making it impossible to cancel the gauge transformation matrices. A common way to maintain the gauge invariance of the action on a lattice is to treat gauge fields as link variables between the sites of the lattice. We describe this below. \\

Consider the action
\begin{equation}
S_E = \frac{N}{2 \lambda} \int_0^{\beta} d t \operatorname{Tr}\left[\left(\mathcal{D}_{t} X^{i}\right)^{2}-\frac{1}{2}\left[X^{i}, X^{j}\right]^{2}\right], \\
\label{eq:2}
\end{equation}
where $\mathcal{D}_{t} X^{i} = \frac{d X^{i}}{d t} - i\left[A_{t}, X^{i}\right]$; $i = 1, \cdots, 9$. This action is invariant under the gauge transformations
\begin{eqnarray}
X^{i}(t) &\longrightarrow& \Omega(t) X^{i}(t) \Omega^{\dagger}(t), \\
A(t) &\longrightarrow& \Omega(t)\left(A(t) + i \frac{d}{d t}\right) \Omega^{\dagger}(t).
\label{eq:3}
\end{eqnarray}

The discrete form of the pure derivative part in $\mathcal{D}_t$ is $\frac{\partial X^i(t)}{\partial t} \xrightarrow{} \frac{X^i_{t+1} - X^i_t}{a}$, which is not invariant under the same transformation. In such cases, link fields are added between the sites of scalar matrices to define the covariant derivative. \\
 
Taking the link fields defined as
\begin{equation}
U_{t, t+1} = \mathcal{P} \exp \left[i \int_{t a}^{(t+1) a} d t A(t)\right],
\end{equation}
with the transformation property
\begin{equation}
    U_{t,t+1} \xrightarrow{} \Omega_t U_{t,t+1}\Omega^{\dagger}_{t+1},
\end{equation}
we can write down the discretized form of covariant derivative as
\begin{equation}
\mathcal{D}_t \rightarrow \frac{1}{a} \left[ U_{t, t+1} X_{t+1}^{i} U_{t+1, t}-X_{t}^{i} \right],
\label{eq:ch1-covdr}
\end{equation}
where $U_{t+1, t} = U_{t, t+1}^{\dagger}$, and $a$ denotes the lattice spacing.

However, in this thesis we will use a slightly different approach. The gauge is chosen such that all the gauge variables interacts only with the scalar matrices at the boundaries of integral and it is diagonalized such that all gauge variables act as angles on the unit circle. This choice of gauge is called the {\it static-diagonal gauge} and we will use this to study the model. Thus we have
\begin{equation}
A_{t} = \operatorname{diag} \left(\theta_{1}, \cdots, \theta_{N}\right).
\end{equation}

With this gauge choice, the link field and the Polyakov loop are written as
\begin{equation}
U = \operatorname{diag}\left(\mathrm{e}^{\mathrm{i} \theta_{1}}, \mathrm{e}^{i \theta_{2}}, \cdots, \mathrm{e}^{i \theta_{N}}\right), 
\end{equation}
and
\begin{equation}
P = \frac{1}{N} \operatorname{Tr} U.
\end{equation}

A new term called the Faddeev-Popov determinant appears as part of the partition function when we use the static-diagonal gauge. (Its origin is due to the change of variables from the $U$ matrices to the angles $\theta_i$.) The partition function with this new term has the form
\begin{equation}
Z \approx \int d \theta_{1} \cdots d \theta_{N} \prod_{l < m}^{N} \sin^{2} \left(\frac{\theta_{l} - \theta_{m}}{2}\right) \int \mathcal{D}[X] e^{-S_{E}[X]}.
\end{equation}

\section{Computing observables using path integrals}

The final task is to put the partition function on a lattice and then to compute the expectation values of observables by evaluating the path integrals. If the partition function has the form $Z = \int \mathcal{D}[X]e^{-S_{E}[X]}$, then expression for the expectation value of an observable is given by
\begin{equation}
\langle\mathcal{O}\rangle = \frac{1}{Z} \int \mathcal{D}[X] e^{-S_{E}[X]} \mathcal{O}[X].
\end{equation}

Integrals of the above form can be calculated using Monte Carlo integration method. This method is based on generation of random numbers to estimate the values of integrals. We will discuss this method in detail in the next chapter.

\chapter{Monte Carlo Techniques}
This chapter introduces the computational techniques required to solve the Euclidean path integrals on the lattice. To begin, consider a system of matrices with the following partition function $Z$ written in terms of a Euclidean action $S_E$.
\begin{equation}
Z = \int \mathcal{D}[X] e^{-S_{E}[X]}.
\end{equation}
The factor $e^{-S_E}$ in the partition function acts as a weighting function for each possible state of the system. 

With this partition function the expectation value of an observable is written as
\begin{equation}
\langle\mathcal{O}\rangle = \frac{\int\mathcal{D}[X] \mathcal{O}[X] e^{\left.-S_{E} [X\right]}}{\int\mathcal{D}[X] e^{-S_{E}[X]}}.
\end{equation}
The integrals in the expectation value of an observable are over all matrix variables at all times. The established method of calculating these integrals is first to break the time variable into smaller slices and then evaluate each spatial integral at a fixed time slice. With this done, the time variable splits into $T$ number of lattice sites with $a$ being the lattice spacing, and the matrix variables defined at all times now exist only at the lattice sites. 

The expressions for the partition function and the expectation value after discretization are
\begin{equation}
Z_{\rm lat} = \left(\prod_{t=1}^{T}\prod_{i=1}^d \int dX_{t}^i\right) e^{-S_{\rm lat}[X]},
\end{equation}
\begin{equation}
\label{eq:obv}
\langle\mathcal{O}\rangle = \frac{\left(\prod_{t=1}^{T}\prod_{i=1}^d \int dX_{t}^i\right) \mathcal{O}[X]e^{\left.-S_{\rm lat} [X\right]}}{\left(\prod_{t=1}^{T}\prod_{i=1}^d \int dX_{t}^i\right) e^{-S_{\rm lat}[X]}},\end{equation}
where $S_{\rm lat}$ represents the discrete form of the Euclidean action. The formalism to discretize the action on a lattice is discussed in the next chapter. The variables $T$ and $a$ have the same meaning throughout this thesis. \\

Now, the problem that is remaining for us is just to solve these path integrals using a suitable numerical algorithm.

\section{Monte Carlo integration method}

The partition function described above contains a large number of integrals even for a small lattice size and a small number of spatial dimensions. The exact evaluation of these integrals would require solving a large number of summations over every possible state of the system, which is clearly impossible. So a new way to estimate these integrals is required. \\

One approach to estimate these integrals comes from the probability theory, which tells that the integral over a function can be approximated by averaging the function over randomly selected points within its domain. This method is called the Monte Carlo integration method. \\

If an integral over function $f(x)$ needs to be evaluated over a domain $Y$, we first select $n$ number of points ($x_i \in Y$) randomly from the domain according to the uniform distribution $\omega_{u}\left(x_{i}\right) =1 /(\beta-\alpha)$. Then an average of the function is calculated over those selected $x_i$ values.
\begin{equation}
\frac{1}{\beta-\alpha} \int_{\alpha}^{\beta} \mathrm{d} x f(x) = \lim _{n \rightarrow \infty} \frac{1}{n} \sum_{i=1}^{n} f\left(x_{i}\right).
\end{equation}

The error of the Monte Carlo integration is $\propto 1/\sqrt{n}$. Thus the accuracy of the integral increases as the number of randomly selected points for the average increases. The exact value of the integral will be reached for $n \xrightarrow{} \infty$. \\

\subsection{Importance sampling}

The expression for the expectation value of an observable in Eq. \eqref{eq:obv} has a Boltzmann weight factor $e^{-S_{\rm lat}}$, which gives different importance to different field configurations. The use of a uniform probability distribution for sampling in such cases results in a poor estimate of the integral. Therefore, it is necessary to consider a different probability distribution, which can be used to sample configurations having large weight factor. Sampling configurations according to their weight is called importance sampling. \\

In importance sampling method, the expectation value of a function $f(x)$ with a probability distribution $\omega(x)$ given by
\begin{equation}
\langle f\rangle_{\omega} = \frac{\int_{\alpha}^{\beta} \mathrm{d} x \omega(x) f(x)}{\int_{\alpha}^{\beta} \mathrm{d} x \omega(x)},
\end{equation}
is approximated as an average over $n$ points,
\begin{equation}
\langle f\rangle_{\omega} = \lim _{n \rightarrow \infty} \frac{1}{n} \sum_{i=1}^{n} f \left(x_{i}\right),
\end{equation}
where each point is randomly sampled according to the normalized probability density
\begin{equation} 
p(x) = \frac{\omega(x) \mathrm{d} x}{\int_{\alpha}^{\beta} \mathrm{d} x \omega(x)}.
\end{equation}

The path integral in Eq. \eqref{eq:obv} is of this form, and therefore, the expectation value of an observable can be obtained by \cite{texbook}
\begin{equation}
\langle O \rangle = \lim _{n \rightarrow \infty} \frac{1}{n} \sum_{i=1}^{n} O \left[X_{[i]}\right],
\label{eq:avg}
\end{equation}
where each of the $X_{[i]}$ is sampled with the probability density 
\begin{equation}
p(X) = \frac{\mathrm{e}^{-S_{\rm lat}[X]} \prod_{t=1}^{T}\prod_{i=1}^d dX_{t}^i}{\prod_{t=1}^{T}\prod_{i=1}^d \int dX_{t}^i \,\mathrm{e}^{-S_{\rm lat}[X]}}.
\label{eq:11}
\end{equation}

\subsection{Markov chains}

The problem now is to produce field configurations with the probability distribution $P(X) \propto \exp(-S_{\rm lat})$. This is done by making use of the Markov process. In a Markov process, we start with some random field configuration and then generate a stochastic sequence of configurations, which ultimately achieve the equilibrium distribution $P(X)$.
\begin{equation}
X_{0} \longrightarrow X_{1} \longrightarrow X_{2} \longrightarrow \ldots
\end{equation}

This sequence of field configurations is called a Markov chain. If $V$ is the volume of states of the system, then the Markov chain is made such as it moves more often in that region of $V$, which corresponds to the configurations having substantial weight $\exp(-S_{\rm lat})$. \\

There are two conditions that a Markov chain must follow to reach the equilibrium distribution. These two conditions are ergodicity and the detailed balance conditions. Ergodicity means that any state of the system should be accessible from any other state. The detailed balance condition is a sufficient condition for proving the invariance (or stationarity) of the probability distribution. The equation given below is called the detailed balance equation
\begin{equation}
T \left(X^{\prime} | X\right) P(X) = T \left(X | X^{\prime}\right) P\left(X^{\prime}\right),
\end{equation}
where $T\left(X^{\prime} | X\right)$ is the transition probability from $X$ to $X^{\prime}$. \\

In the following next two sections, we describe two different algorithms. They both are based on the Markov chain method.

\subsection{Random-walk Metropolis algorithm}

The Metropolis algorithm \cite{doi:10.1063/1.1699114} creates a new field configuration from the previous field configuration using the following steps:
\begin{enumerate}
    \item Given a matrix variable $X_{\rm old}$ at time $t$. \\
    \item Create a proposed matrix by adding a small random matrix to the old matrix variable, i.e. $X^{\prime} = X_{\rm old} + \Delta X$. \\
    \item Use the Metropolis test to accept or reject the proposed matrix. \\
    In this, the proposed matrix is accepted with a probability \\
    \begin{equation} 
    {\rm min} \left(1, e^{-\Delta S}\right).
    \end{equation}
    We apply Step $3$ in the code as:
    \begin{equation} 
    X_{\rm new} = \left\{\begin{array}{cl}
X^{\prime} & \Delta S\leq0 \\ \\
X^{\prime} \text{ if }r\leq e^{-\Delta S} & \Delta S > 0\\ X_{\rm old} \text{ if } r > e^{-\Delta S},
\end{array}\right.
\end{equation}
where $r$ is a uniform (pseudo)random number in interval $(0,1)$, and $\Delta S = S^{\prime}_{\rm lat} - S_{\rm lat}$.
    \item Repeat the steps 2 and 3 to create a sequence of configuration.
\end{enumerate}

There are various algorithms to generate pseudo-random numbers in the interval $(0,1)$. Some of these methods are given in Ref. \cite{PresTeukVettFlan92}. These numbers are called pseudo-random numbers because they pass most of the tests of randomness even though an algorithm produces them.

\subsection{Hamiltonian Monte Carlo (HMC) algorithm}

Hamiltonian Monte Carlo (HMC) \cite{DUANE1987216} \cite{neal2012mcmc} is another method that constructs the Markov chain by using the Metropolis test. It uses Hamilton's equations for this purpose. \\


We provide an overview of this methodology below. \\

For simplicity, consider a zero-dimensional matrix model with a Euclidean action, $S_{E}[X]$. Let us assume that $S_{E}[X]$ is a function of an $N \times N$ scalar Hermitian matrix.\\ 

First step of this method is to think the scalar matrix as a function of a fictitious time, $\tau$.  
\begin{equation}
X \equiv X(\tau).
\end{equation}

Second step is to define a Hamiltonian function for the system. For this, a kinetic energy term constituting of an $N \times N$ Hermitian momentum matrix is added to the action. The action, here, behaves as a potential energy. The Hamiltonian of the system is then written as
\begin{equation}
H = \frac{1}{2} \operatorname{Tr} {P}^{2} + S_{\mathrm{E}}[X].
\label{eq:ham}
\end{equation}

Further, the equations of motion are calculated using the Hamilton's equations. The Hamilton's equations are given by 
\begin{equation}
\frac{\partial H}{\partial\left(P\right)_{a b}} = \left(\dot{X}\right)_{a b}, ~~~ \frac{\partial H}{\partial\left(X\right)_{a b}} = -\left(\dot{P}\right)_{a b}
\end{equation}
      
Then the equations of motion are  
\begin{equation}
\left(P\right)_{b a} = \left(\dot{X}\right)_{a b},
\label{eq:em1}
\end{equation}

\begin{equation}
\frac{\partial S_{\mathrm{E}}}{\partial\left(X\right)_{a b}} = -\left(\dot{P}\right)_{a b}, 
\label{eq:em2}
\end{equation}
where the dots indicate that the derivatives are with respect to $\tau$.

\subsubsection{The leapfrog method}

Next step is to solve the differential equations in Eqs. \eqref{eq:em1} and \eqref{eq:em2}. The following set of equations is used for this purpose.
\begin{equation}
\left(P\right)_{a b}\left(\tau + \frac{ \epsilon}{2}\right) = \left(P\right)_{a b}(\tau) - \frac{\epsilon}{2} \frac{\partial S_{E}}{\partial\left(X\right)_{a b}(\tau)},
\end{equation}
\begin{equation}
\left(X\right)_{a b}(\tau+ \epsilon) = \left(X\right)_{a b}(\tau) + \epsilon\left(P\right)_{b a}\left(\tau + \frac{ \epsilon}{2}\right),
\end{equation}
\begin{equation}
\left(P\right)_{a b}(\tau + \epsilon) = \left(P\right)_{a b}\left(\tau + \frac{ \epsilon}{2}\right) - \frac{ \epsilon}{2} \frac{\partial S_{E}}{\partial\left(X\right)_{a b}(\tau + \epsilon)},
\end{equation}
where $\epsilon$ represents the time step, in fictitious simulation time, on the lattice. The above set of equations form part of the leapfrog algorithm. \\

Thereafter a Metropolis test is performed to accept or reject the proposed state.

\subsubsection{Properties of Hamiltonian dynamics}

Though Hamiltonian dynamics has many properties, its three properties are significant for its use in constructing the Markov chain Monte Carlo updates.
\begin{enumerate}
    \item Hamiltonian dynamics is reversible. This property is can be used to show that the MCMC (Markov chain Monte Carlo) updates, which use the Hamiltonian dynamics leaves the desired probability distribution invariant.
    \item Hamiltonian dynamics keeps the Hamiltonian invariant, i.e., conserved. But practically, in numerical simulations, the Hamiltonian can only be made approximately invariant.
    \item Hamiltonian dynamics keeps the volume of the space preserved. This property ensures that no additional term is needed to add in the acceptance probability for the Metropolis update.
\end{enumerate}

\subsubsection{The HMC algorithm}

The steps involved in HMC algorithm are provided below.\\
Consider a lattice of points $\tau = n \epsilon$, where $n=0, \cdots,\nu-1, \nu$. 
\begin{enumerate}
    \item Given $X$ = $X(0)$.
    \item Construct $P = P(0)$ according to the Gaussian distribution, $e^{-\frac{1}{2} {\rm Tr} {P}^2}$. Assign $X$ the $P$ matrix.
    \item Solve the differential equations using the Leapfrog method to get the configuration $({X}^{\prime},{P}^{\prime})\equiv(X(\sigma),P(\sigma))$. Here $\sigma=\nu \epsilon$.
    \item Use the Metropolis test to accept or reject the configuration $({X}^{\prime},{P}^{\prime})$. The configuration is accepted with a probability 
    \begin{equation} {\rm min} \left(1, e^{-\Delta H(X,P)}\right),
    \end{equation} 
    where $\Delta H = H(X^{\prime}, P^{\prime}) - H(X, P)$
    \item Repeat the steps 2 to 4.
\end{enumerate}

At the beginning of each molecular dynamics trajectory, the old momentum matrix is dropped, and a new matrix is constructed. In this way we maintain ergodicity of the system. \\

The use of the Metropolis test confirms the detailed balance of this algorithm, and also the absence of systematic errors occurred due to the non-conservation of the discrete Hamiltonian.

\subsubsection{Worked example: a 1-D system}

Consider a 1-D system with the following Hamiltonian.

\begin{equation}
H(q, p) = V(q) + K(p),
\end{equation}
where
\begin{equation}
V(q) = q^{2} / 2, \quad K(p) = p^{2} / 2,
\end{equation}
with $p$ and $q$ being scalars. \\

The equations of motion are
\begin{equation}
\frac{d q}{d t} = p, \quad \frac{d p}{d t} = - q.
\end{equation}

Solutions of these differential equations have the form
\begin{equation}
q(t) = \beta \cos (\alpha + t), \quad p(t) = -\beta \sin (\alpha + t).
\end{equation}

Thus the trajectory in the phase diagram of $q$ vs $p$ is circular.

\begin{figure}[ht!]
\begin{center}
\includegraphics[scale=0.6]{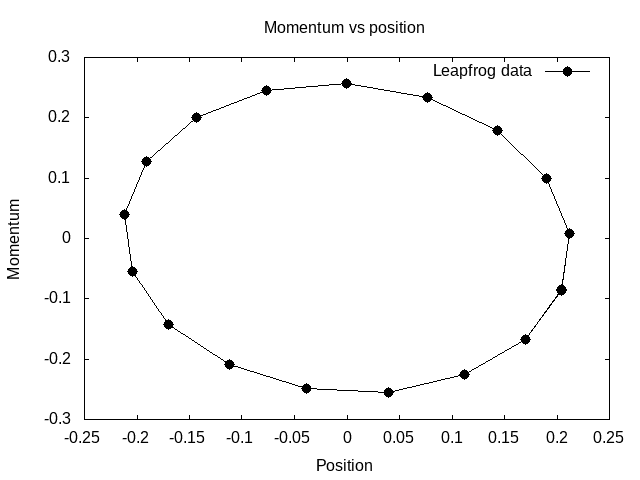}
\end{center}
\caption[The plot of $q$ vs $p$ using HMC algorithm with step size $\Delta \epsilon = 0.3$.]{The plot of position ($q$) vs momentum ($p$) using HMC algorithm for with step size $\Delta \epsilon = 0.3$.}
\label{fig:2}
\end{figure}

\begin{figure}[ht!]
\begin{center}
\includegraphics[scale=0.6]{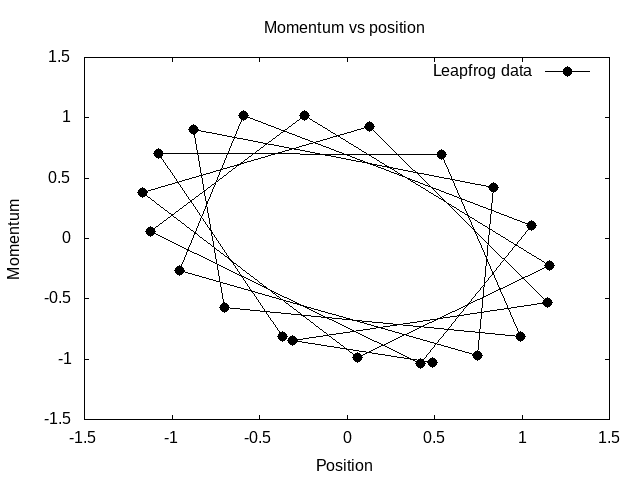}
\end{center}
\caption[The plot of $q$ vs $p$ using HMC algorithm with step size $\Delta \epsilon = 1.2$.]{The plot of position ($q$) vs momentum ($p$) using HMC algorithm with step size $\Delta \epsilon = 1.2$.}
\label{fig:3}
\end{figure}

In Figs. \ref{fig:2} and \ref{fig:3} we show the trajectories obtained after the application of the HMC algorithm to this system.


\section{Statistical error and autocorrelation}

The following formula can be used to calculate the statistical error in the data sampled
\begin{equation}
\delta O = \frac{\sigma}{\sqrt{n}},
\end{equation}
where
\begin{equation}
\sigma^{2} = \left \langle O^{2} \right \rangle - \langle O \rangle^{2}.
\end{equation}

However, the formula is valid only for the case when configurations obtained after the thermalization (explained below) are uncorrelated, i.e., independent. In real simulations, two consecutive configurations are dependent, and using such configurations for calculating the average results in a skewed value. It is possible to obtain an average number of sweeps, which needs to be skipped to get two uncorrelated configurations. This is done by calculating the function in Eq. \eqref{eq:auto}, and then evaluating the point where it becomes zero. The average number of sweeps separating two consecutive uncorrelated configurations is called autocorrelation length. Taking $j$ as a positive integer called the {\it lag time}, we have the {\it lag-$j$ autocovariance function}
\begin{equation}
\label{eq:auto}
\Gamma_j = \frac{1}{n-j} \sum_{i=1}^{n-j} \left(O_{i} - \langle O \rangle \right) \left(O_{i+j} - \langle O \rangle \right).
\end{equation}

Normalizing the above expression we get the autocorrelation function is
\begin{equation}
\rho_j = \frac{\Gamma_j }{\Gamma_0},
\end{equation}
where $\Gamma_0 = \sigma^2$.

\section{Advantages of HMC algorithm}

There are two important advantages of HMC, which make it better than the random-walk Metropolis algorithm. They are listed below: 
\begin{enumerate}
    \item HMC converges to the equilibrium distribution faster than the other algorithm.
    \item It has very low autocorrelation time as compared to the other.
\end{enumerate}

Actually, in the random-walk Metropolis algorithm, a Markov chain makes jumps in state space at random directions due to which it takes a long time to converge. Whereas, in HMC, the addition of auxiliary variables in action, guides a Markov chain to make jumps in `more appropriate' directions.\\   

Because of these two benefits it is possible to save a lot of computing time.

\section{Simulation steps}

For simulations of the models discussed in this thesis, the matrix variables are placed on a one-dimensional lattice.
\begin{enumerate}
\item Simulations can be started either with a cold start or a hot start depending upon the system. If the initial matrix variables are null matrices, then we call it a cold start. If they are random, we call it a hot start.
\item Then a Markov chain of configurations is constructed using any of the two algorithms. The proposed matrix variable is accepted or rejected according to the Metropolis test. If accepted, the old matrix variable is replaced by the proposed matrix variable, otherwise the old matrix variable is kept as a new matrix variable. This process is called an update.
\item The system is then let to evolve for enough number of sweeps until it reaches the equilibrium distribution. This process is called thermalization. A sweep involves the update of entire lattice at once.
\item After the system gets thermalized, average of an observable can be calculated using Eq. \eqref{eq:avg}. The data points for this average should be sampled with an appropriate autocorrelation length to minimize the statistical error.
\end{enumerate}

\chapter{Harmonic Oscillator}
A good model to apply the simulation methods we encountered in the last chapter is the harmonic oscillator. The action of this model is given by
\begin{equation}
S_E = \frac{1}{2} \int_{0}^{\beta} d t \operatorname{tr}\left[\dot{X}^{2} + \mu^{2} X^{2}\right].
\end{equation}
Here $\beta$ is defined as inverse of the temperature, $\mu$ is a mass parameter and the $X$s are scalar $N \times N$ Hermitian matrices with periodic boundary conditions $X(t + \beta) = X(t)$. The `gauge field' is not included here; we will discuss a model containing the gauge field in the next chapter. \\

To apply the techniques discussed in the last chapter, the action must be discretized first. In order to do so, the integral with respect to time is substituted for summations, and the derivative is substituted for finite difference operators. We have
\begin{equation}
\begin{aligned}
\int_{0}^{\beta} d t & \sim a \sum_{t=1}^{T=\frac{\beta}{a}}, \\
\frac{\partial X_t^i}{\partial t} & \sim \frac{X_{t+1}^i - X_{t}^i}{a}, \\
X^i(t) & \sim X^i_{t}. \\
\end{aligned}
\end{equation}

With these changes, the lattice action is given by
\begin{equation}
\label{eq:har1}
S_{\rm lat} = \frac{a}{2} \sum_{t=1}^{T} \operatorname{tr} \left[\left(\frac{X_{t+1} - X_{t}}{a}\right)^{2} + \mu^{2} X_{t}^{2}\right].
\end{equation}

The next step is to generate the field configurations with weight  $e^{-S_{\rm lat}}$.

\section{Random-walk Metropolis for the model}

In this section, a procedure to apply the Metropolis algorithm is outlined. The method is first to add a random matrix to one scalar matrix at a time and then check the effect of the change. By looking at Eq. \eqref{eq:har1}, it can be inferred that evaluating a local action is advantageous. The local action includes the terms that consist of the scalar matrix to which the change is made. The local action of the model is given by 
\begin{equation}
S_{\rm loc} = \frac{a}{2} \operatorname{tr} \left[\left(\frac{2}{a^{2}} + \mu^{2}\right) \left(X_{t}\right)^{2} - 2 X_{t}\left(\frac{X_{t+1} + X_{t-1}}{a^{2}} \right) \right].
\end{equation}

This action reflects the total change in the full action that occurs from altering the scalar matrix at a given lattice site.\\

The addition of a random matrix to a scalar matrix can be shown as
\begin{equation}
\label{eq:har2}
    X^{\prime} = {X^{\rm old}}_t + \delta \Omega.
\end{equation}
Here $\Omega$ is a Hermitian matrix filled with real and complex random numbers.\\

\noindent In the Metropolis test, the proposed change is accepted if 
\begin{equation}
r \leq \exp \left[-S_{\rm loc} \left(X_{t}^{\prime}\right) + S_{\rm loc} \left(X_{t}^{\rm old}\right) \right],
\end{equation}
otherwise it is rejected. Here $r \in(0,1)$ is a uniform deviate. Before moving to the next lattice site, the given site should be updated numerous times to achieve the thermalization faster. This entire process is repeated for every lattice site until the whole lattice is covered. A sweep is completed when entire lattice has been updated. \\

\begin{figure}[ht!]
\begin{center}
\includegraphics[scale=1.0]{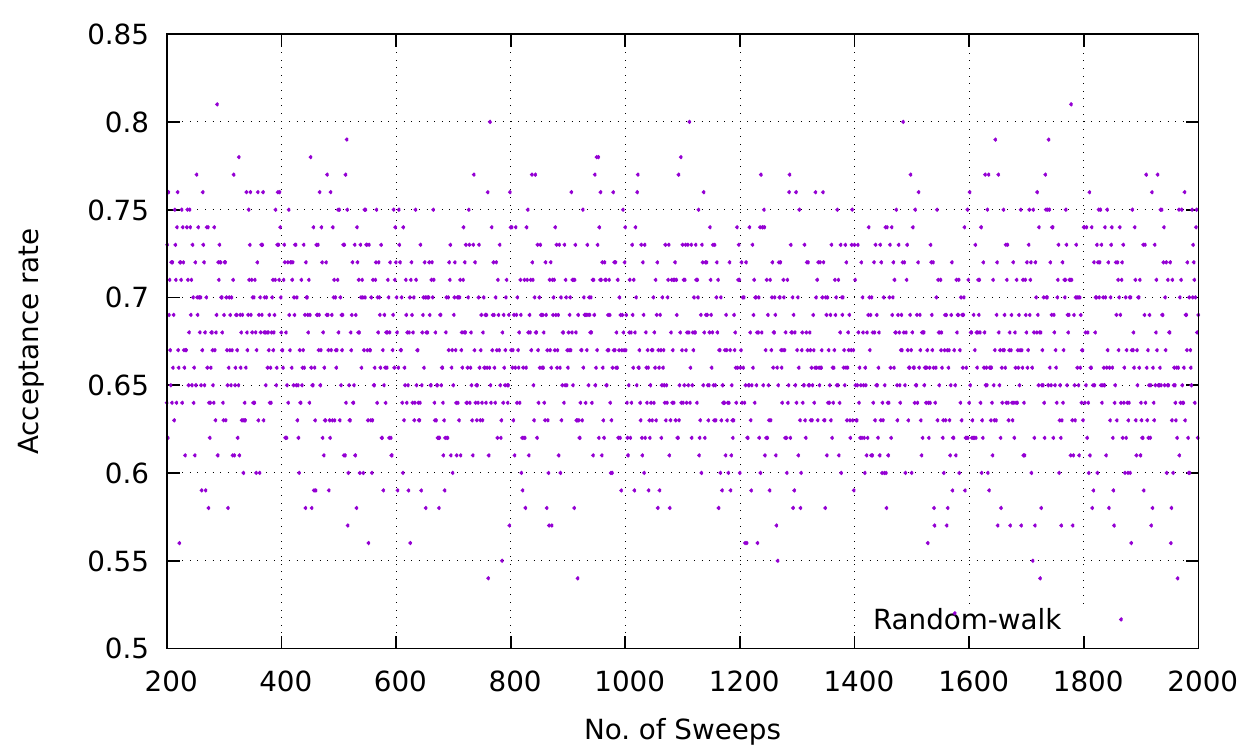}
\end{center}
\caption[Acceptance rate for random-walk Metropolis.]{A plot of the acceptance rate for the harmonic oscillator model using random-walk Metropolis algorithm. Here, $N = 16$, $d = 1$, $\mu = 1$, $T = 32$, $a = 0.3$ and $\delta = 0.0250$. This value of $\delta$ stands for the case where each lattice site has been updated 25 times in one sweep.}
\label{fig:ch4-acceptance_rate}
\end{figure}

In Eq. \eqref{eq:har2}, a new parameter $\delta$ has been introduced to control the acceptance rate. An acceptance rate is defined as the ratio of accepted updates to the total number of proposed updates. The value for the parameter $\delta$ is chosen such that the acceptance rate lie in the range $60\%$ to $75\%$. (See Fig. \ref{fig:ch4-acceptance_rate}.)

The reasons for this range are as follows
\begin{itemize}
    \item If the acceptance rate is kept too low, then the Markov chain would make longer jumps in configuration space due to which it would skip major regions of importance.
    \item If the acceptance rate is kept too high, then the Markov chain would make shorter jumps in configuration space due to which it would not be able to cover the whole configuration space.
\end{itemize}

\subsection{\texorpdfstring{Construction of $\Omega$ matrix}{}}

The following procedure can be used to create elements of the matrix. \\

For off-diagonal elements
\begin{equation}
\begin{aligned}
&\left(\Omega\right)_{a b} = n_{1} + i n_{2}, \\
&\left(\Omega\right)_{b a} = \left(\Omega\right)^{*}_{a b}.
\end{aligned}
\end{equation}

For diagonal elements
\begin{equation}
\begin{aligned}
&\left(\Omega\right)_{a a} = n_{1}.
\end{aligned}
\end{equation}
Here, $n_1$ and $n_2$ are uniform random deviates in $(-1, 1)$. 

\subsection{Periodicity on the lattice}

Since the scalar matrices are periodic, Eq. \eqref{eq:ch4-period} can be used to manage this condition on the lattice. We have
\begin{equation}
\label{eq:ch4-period}
    X_{T+t} = X_t,
\end{equation}
where $t = 1, \cdots, T$.

\section{HMC for the model}

This section explains how HMC algorithm can be applied to the model. The procedure consists of multiple steps which are as follows
\begin{enumerate}
    \item Construct a momentum matrix for a given lattice site.
    \item Use this matrix to create a local Hamiltonian.
    \item Get the equations of motion for that lattice site.
    \item Solve these equations to get a proposed scalar matrix and a momentum matrix for that site.
    \item Use the Metropolis test to update the site.
    \item Then move to another site and repeat the same process.
\end{enumerate}

\noindent It is beneficial to calculate a local Hamiltonian for this method also. \\ 

\noindent The local Hamiltonian for this model is given by 
\begin{equation}
H_{\rm loc} = \frac{1}{2} \operatorname{tr} P_t^2 + \frac{a}{2} \operatorname{tr} \left[\left(\frac{2}{a^{2}} + \mu^{2}\right) \left(X_{t}\right)^{2} - 2 X_{t} \left(\frac{X_{t+1} + X_{t-1}}{a^{2}} \right) \right].\end{equation}
Here, $P_t$ is a Hermitian momentum matrix with real and complex random numbers as its elements, and it is constructed according to the Gaussian distribution $\exp \left( - \frac{1}{2} \operatorname{tr} P_t^2 \right)$. \\

The equations of motion for $H_{\rm loc}$ are 
\begin{equation}
     ( \dot{X_t} )_{a b} = (P_t)_{b a},
\end{equation}

\begin{equation}
    -( \dot{P_t} )_{a b} = a \left[ \left(\frac{2}{a^2} + {\mu}^2 \right)(X_t)_{b a} - \left(\frac{X_{t+1} + X_{t-1}}{a^2}\right)_{b a} \right].
\end{equation}

These differential equations are solved using the leapfrog method.

The steps involved in the leapfrog algorithm are the following.
\begin{itemize}
    \item At a given simulation time $\tau$, firstly, a half-step time evolution ($\epsilon/2$) for the momentum matrix is performed, and then a full-step time evolution ($\epsilon$) for the scalar matrix is made using the new momentum matrix. Finally, a half-step time evolution ($\epsilon/2$) for the momentum matrix is performed again by using the new scalar matrix. These time steps are repeated multiple times to get a distant configuration, and this is done by assuming a trajectory of a fictitious time variable over which iterations are carried out by using an appropriate leapfrog time step $\epsilon$. The steps involved in the leapfrog process are summarized in the following equations.
\end{itemize}

\begin{equation}
\left(P_t\right)_{a b} \left(\tau + \frac{\epsilon}{2}\right) = \left(P_t\right)_{a b}(\tau) - \frac{\epsilon}{2} \frac{\partial S_{\rm loc}}{\partial\left(X_{t}\right)_{a b}(\tau)},
\end{equation}
\begin{equation}
\left(X_{t}\right)_{a b}(\tau + \epsilon) = \left(X_{t}\right)_{a b}(\tau) + \epsilon\left(P_{t}\right)_{b a}\left(\tau + \frac{ \epsilon}{2}\right). 
\end{equation}
\begin{equation} 
\left(P_{t}\right)_{a b}(\tau + \epsilon) = \left(P_{t}\right)_{a b}\left(\tau + \frac{ \epsilon}{2}\right) - \frac{ \epsilon}{2} \frac{\partial S_{\rm loc}}{\partial\left(X_{t}\right)_{a b}(\tau + \epsilon)},
\end{equation}
where $\tau$ represents fictitious time, $\epsilon$ represents lattice spacing of the lattice made from $\tau$ and 
\begin{equation}
\frac{\partial S_{\rm loc}}{\partial\left(X_{t}\right)_{a b}(\tau)} = a \left[ \left(\frac{2}{a^2} + {\mu}^2 \right)(X_t)_{b a}(\tau) - \left(\frac{X_{t+1} + X_{t-1}}{a^2}\right)_{b a}\right].
\end{equation}

\noindent In the Metropolis test, the solutions obtained after the application of the leapfrog method are accepted if
\begin{equation}
r \leq \exp \left[ -H_{\rm loc} \left( P_t^{\prime}, X_t^{\prime} \right) + H_{\rm loc} \left( P_t^{\rm old}, X_t^{\rm old} \right) \right],
\end{equation}
otherwise they are rejected. When the given lattice is updated once, the updated momentum matrix is discarded, and a new momentum matrix is constructed for the next site. The parameter $\epsilon$ serves the same purpose of controlling the acceptance rate as the parameter $\delta$ does in the random-walk Metropolis method. The value for the parameter $\epsilon$ is chosen by observing the acceptance rate and the autocorrelation function on trial runs. There is no need to perform multiple updates within a sweep for this method. \\

\begin{figure}[ht!]
\begin{center}
\includegraphics[scale=1.0]{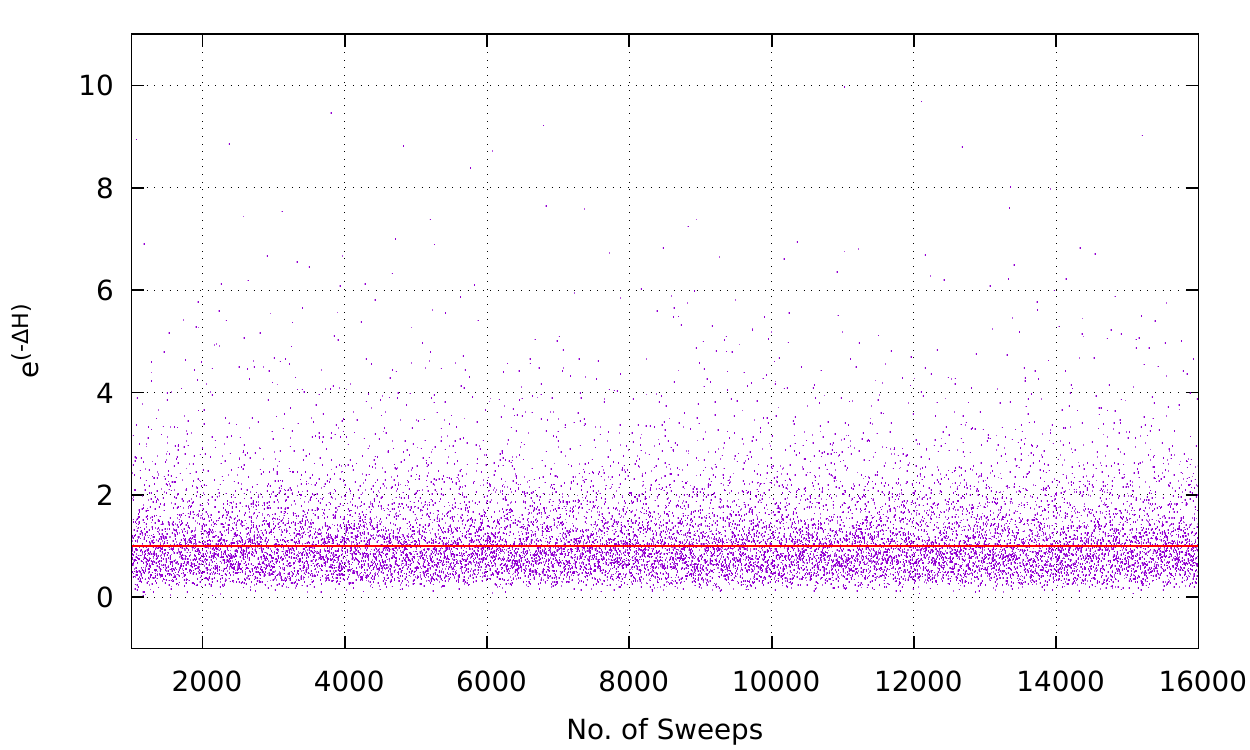}
\end{center}
\caption[The plot of $\exp (- \Delta H)$ against the number of sweeps.]{The plot of $\exp (-\Delta H)$ against the number of sweeps. Here, $\Delta H = H_{\rm loc}(P_{t}^{\prime}, X_{t}^{\prime}) - H_{\rm loc}(P_{t}^{\rm old}, X_{t}^{\rm old})$. The horizontal solid line represents $\exp(- \Delta H) = 1$.}
\label{fig:delH_ch4}
\end{figure}

The correctness of the gradient of action can be checked by using $e^{-\Delta H}$ shown in Fig. \ref{fig:delH_ch4}. If there is no sign error or any term missing in the gradient of action then the values of $e^{-\Delta H}$ would lie around 1. However, rare random points may jump beyond it as proposed states are accepted with some errors.

\subsection{\texorpdfstring{Construction of $P_{t}$ matrix}{}}

The following technique can be applied to create the elements of the $P_t$ matrix. \\

For off-diagonal elements we take
\begin{equation}
\begin{aligned}
&\left(P_{t}\right)_{a b} = \frac{g_1 + i g_2}{\sqrt{2}}, \\
&\left(P_{t}\right)_{b a} = \left(P_{t}\right)^{*}_{a b}.
\end{aligned}
\end{equation}

For diagonal elements we have
\begin{equation}
\begin{aligned}
&\left(P_{t}\right)_{a a} = g_1.
\end{aligned}
\end{equation}
Here,  $g_1$ and $g_2$ are random deviates generated with a normal (Gaussian) distribution 
\begin{equation}
p(y) d y = \frac{1}{\sqrt{2 \pi}} e^{-y^{2} / 2} d y.
\end{equation}

An algorithm to generate random deviates with this distribution is given in Ref. \cite{PresTeukVettFlan92}.

\section{Comparison between Random-walk Metropolis and HMC}

\subsection{Run-time history}

\begin{figure}[ht!]
\begin{center}
\includegraphics[scale=1.0]{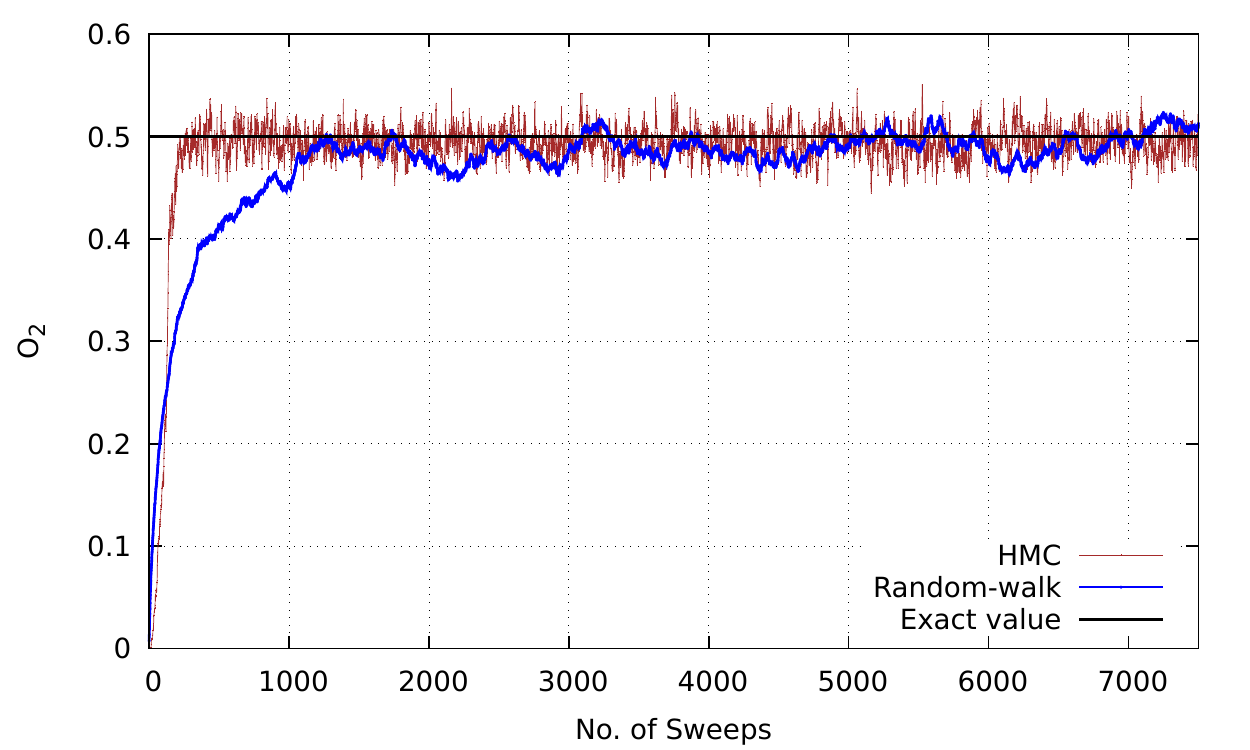}
\end{center}
\caption[Run-time histories of HMC and random-walk Metropolis.]{The run-time history of observable $O_2$ for HMC and random-walk Metropolis. The solid horizontal line is the exact value. Here, $N = 16$, $d = 1$, $\mu = 1$, $T = 32$, $a = 0.3$ and $\beta=aT$.}
\label{fig:ch4-runtimehistory}
\end{figure}

A run-time history of an observable is the behavioral details of the observable from the beginning of the simulation. The observable in Eq. \eqref{eq:ch4-1} has been calculated for the model using both the algorithms and its run-time history is shown in Fig. \ref{fig:ch4-runtimehistory}.
\begin{equation}
\label{eq:ch4-1}
    O_2 = \frac{1}{N^2 T} \sum_{t=1}^{T} \operatorname{tr}(X_t)^{2}.
\end{equation}
In both cases, all the matrix fields have been set to zero at the start of the simulation and then they are left to be evolved. The value of the observable begins from zero, then reaches the theoretical value and oscillates around it. However, it can be seen from Fig. \ref{fig:ch4-runtimehistory} that the observable thermalized much earlier in the case of HMC compared to the random-walk Metropolis.

\subsection{Autocorrelation}

It is evident from Fig. \ref{fig:ch4-runtimehistory} that the value of the observable after each sweep is not independent, rather it is related to the values obtained in the preceding sweeps. It requires multiple sweeps before the observable becomes independent from its previous values. The autocorrelation function defined in Eq. \eqref{eq:auto} is used to determine the average number of sweeps to be skipped for obtaining two uncorrelated values of an observable. \\

\begin{figure}[ht!]
\begin{center}
\includegraphics[scale=1.0]{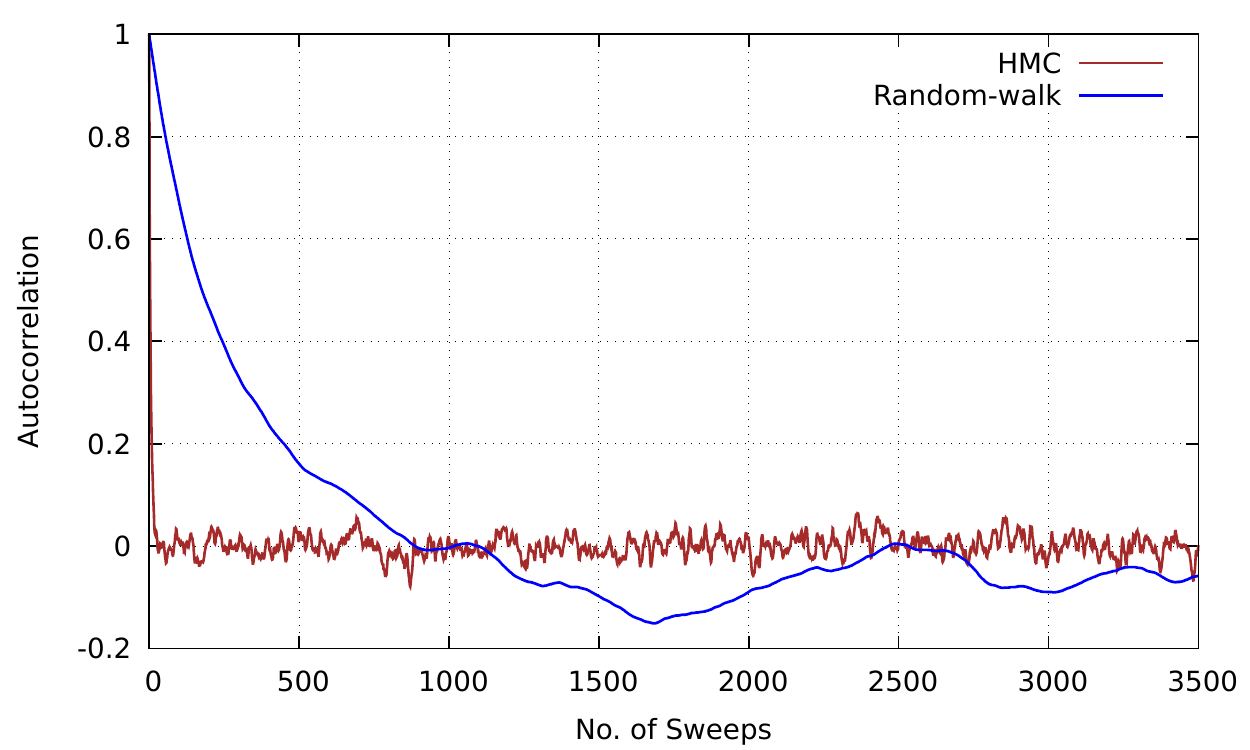}
\end{center}
\caption[Comparison of autocorrelations for random-walk Metropolis and HMC.]{Comparison of autocorrelations for random-walk Metropolis and HMC for observable $O_2$.}
\label{fig:ch4-auto}
\end{figure}

Figure \ref{fig:ch4-auto} shows the plot of normalized autocorrelation for both the methods. In the case of random-walk Metropolis method, the autocorrelation curve begins from one and then slowly drops off to zero in an exponential fashion. It crosses zero nearly around sweep number 900. In HMC, the autocorrelation curve begins from the same value and falls off to zero more quickly. It crosses zero after about sweep number 28, and then continues to fluctuate about it.

\section{Two-point correlation function}

Another interesting observable we can examine in this model is the two-point correlation function. This function is the correlation between matrices at different lattice sites. The value of this function can be calculated using the following formula
\begin{equation}
    O_3 = \left \langle \frac{1}{N^2 } \operatorname{tr} \left[X_{0} X_{t} \right] \right \rangle = \frac{e^{-\mu t} + e^{-\mu(\beta-t)}}{2 \mu\left(1 - e^{-\beta \mu}\right)}.
\end{equation}

\begin{figure}[hbt!]
\begin{center}
\includegraphics[scale=1.0]{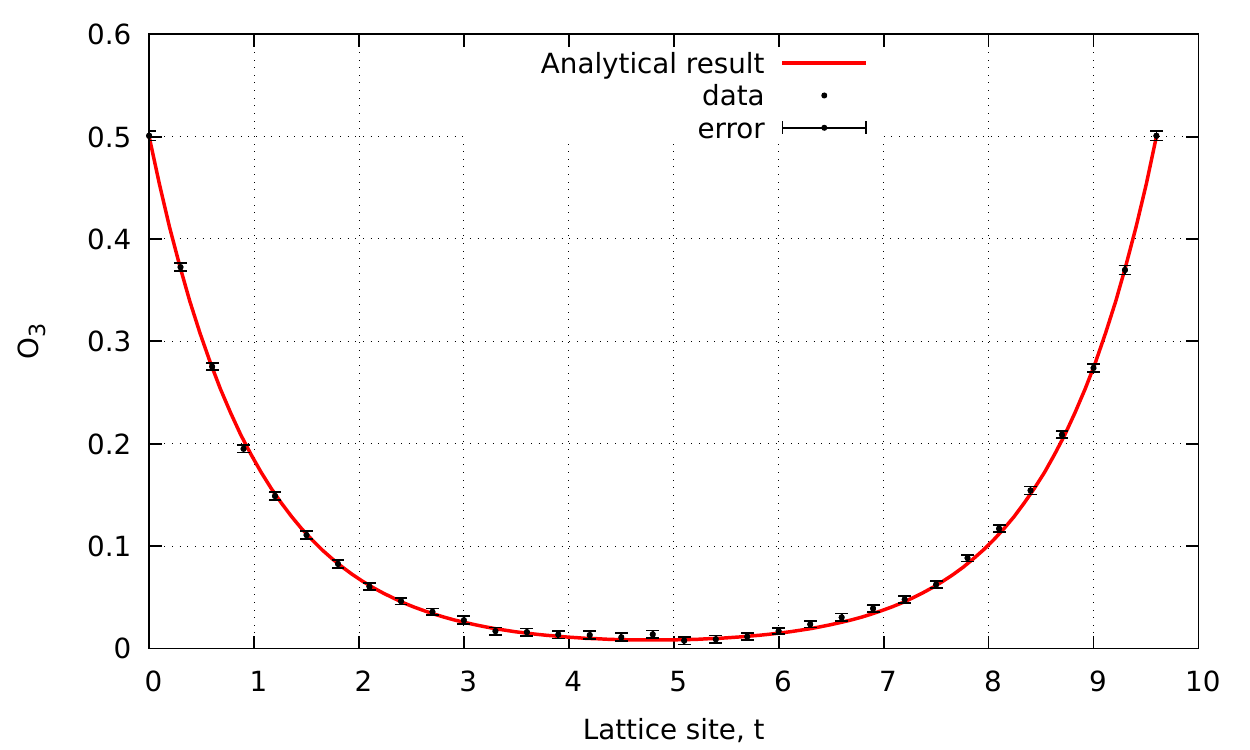}
\end{center}
\caption[Two-point correlation function for harmonic oscillator.]{The two-point correlation function for the harmonic oscillator. Here, $N = 16$, $d = 1$, $\mu = 1$, $T = 32$ and $a = 0.3$. We used HMC algorithm to generate the simulation data.}
\label{fig:ch4-2point}
\end{figure}

Fig. \ref{fig:ch4-2point} shows the two-point correlation function for this model. The plot shows that the simulated data is in good agreement with the theoretical values.

\section{Inference}

We see that the HMC algorithm is relatively faster, and it takes lesser time to generate thermalized uncorrelated configurations, whereas the random-walk Metropolis algorithm is slower and takes more time to generate such configurations. Hence HMC is more useful to simulate more complex models such as the BMN matrix model.

\chapter{Harmonic Oscillator with Commutator Potential}
Now that the basic steps to perform the lattice simulations of the model under question has been established, we can move towards simulating another interesting model. The second model is the harmonic oscillator with a commutator potential. This model includes a commutator squared potential instead of the mass potential, and it carries multiple scalar fields (generated as a result of the compactification of spatial dimensions of a mother theory). 

The Euclidean action for this model is given by
\begin{equation}
    S_E = \frac{N}{2\lambda} \int_{0}^{\beta} d t \operatorname{tr}\left[{(\dot{X^i}})^{2} - \sum_{i<j}^{d}\left[X^{i}, X^{j}\right]^{2}\right].
\end{equation}
Here, $i, j = 1,\cdots, d$ and $\lambda = g_{\rm YM}^2 N$ is the `t Hooft coupling. The partition function is written as $Z = \int \mathcal{D}[X]e^{-S_E}$. \\

Using the methodology explained previously, the lattice action is written as 
\begin{equation}
    S_{\rm lat} = \frac{N a}{2\lambda} \sum_{t=1}^{T} \operatorname{tr} \left[\sum_{i=1}^{d} \left(\frac{X_{t+1}^{i} - X_{t}^{i}}{a}\right)^{2} - \sum_{i<j}^{d} \left[X_{t}^{i}, X_{t}^{j}\right]^{2}\right].
\end{equation}

This action is invariant under the transformation 
\begin{equation}
X_{t}^{i} \xrightarrow{} X_{t}^{i} + \alpha^i \textbf{1},
\end{equation}
where $\alpha^i$ is an arbitrary constant.

In order to remove the corresponding zero mode, the below condition is used.
\begin{equation}
     \sum_{t=1}^{T}\operatorname{tr}(X_{t}^i) = 0 \quad \text{for each $i$.}
\end{equation}

This condition has to be used whenever a commutator potential is present in the action.

\section{HMC for the model}

The same steps, as mentioned in the previous chapter, are applicable here for the use of HMC. These steps must be repeated for updating the matrices of each spatial dimension. The local Hamiltonian is used again for the Metropolis test to update the sites. \\

The local action for site $t$ (with $i$ fixed) is
\begin{equation}
    S_{\rm loc} = \frac{N a}{2\lambda} \operatorname{tr} \left[\frac{2}{a^{2}}\left(X^i_{t}\right)^{2} - 2 X^i_{t}\left(\frac{X^i_{t+1} + X^i_{t-1}}{a^{2}}\right) - \sum_{j\neq i}^{d} \left[X_{t}^{i}, X_{t}^{j}\right]^{2}\right].
\end{equation}

The commutator terms contributing to local interactions are only added in the local action. \\

Then local Hamiltonian and gradient of the action have the following forms
\begin{equation}
    H_{\rm loc} = \frac{1}{2} \operatorname{tr}{P^i_t}^2 + \frac{N a}{2\lambda} \operatorname{tr} \left[\frac{2}{a^{2}}\left(X^i_{t}\right)^{2} - 2 X^i_{t}\left(\frac{X^i_{t+1} + X^i_{t-1}}{a^{2}}\right) -  \sum_{j \neq i}^{d} \left[X_{t}^{i}, X_{t}^{j}\right]^{2}\right],
\end{equation}

\begin{equation}
    \frac{\partial S_{\rm loc}}{\partial (X^i_t)_{a b}} = \frac{N a}{\lambda} \left\{ \frac{2}{a^2}(X^i_t)_{b a} - \left(\frac{X^i_{t+1} + X^i_{t-1}}{a^2}\right)_{b a} - \sum_{j\neq i}^d\left[X^j_t,\left[X^i_t,X^j_t\right]\right]_{b a} \right\}.
\end{equation}

\subsection{The observable $O_4$}

The following observable has been calculated for this model.

\begin{equation}
\label{eq:ch5-o4}
O_4 = \frac{1}{N T} \sum_{t=1}^{T} \sum_{i=1}^d \operatorname{tr} \left(X_{t}^{i}\right)^{2}.
\end{equation}

It is similar to the observable $O_2$ from the previous chapter, except that it includes an extra sum, which is over the spatial dimensions, and has one less power over $N$, which is due to the factor $N/\lambda$ in the action.

\section{Taming the flat directions}

This model has flat directions: they occur due to indefinitely increasing eigenvalues of commutating matrices. Flat directions may encounter at any temperature, but generally, they appear in simulations which are carried out at low temperatures. As long as flat directions persist, the partition function diverges, and Monte Carlo simulations become unstable and eventually break down. So, a method is required to eliminate the problem of flat directions. One way is to add a mass term into the action. The addition of a mass term restricts the eigenvalues of commuting matrices to a finite distribution. This in turn lifts the flat directions and provides stability to simulations. A procedure to calculate the value of observable $O_4$ (see in Eq. \eqref{eq:ch5-o4}) using this technique is given below. \\

\subsection{Simulation procedure}

First, a mass parameter, $\mu$, is introduced in the action by the addition of the term
\begin{equation}
\label{eq:ch5-mterm}
\frac{N}{2\lambda} \int dt \operatorname{tr} \left[\sum_i^d\mu^2 {X^i}^2\right].
\end{equation}
Then the observable is calculated for different small values of $\mu$ to create a plot between the average value of the observable and $\mu$. The obtained plot is fitted with a line or any other suitable function, which can then be extrapolated to get the value at zero mass. \\

With the term given in Eq. \eqref{eq:ch5-mterm}, the expressions for local Hamiltonian and gradient of action are given by 
\begin{equation}
    H_{\rm loc} = \frac{1}{2} \operatorname{tr}{P^i_t}^2 + \frac{N a}{2\lambda} \operatorname{tr} \left[\frac{2}{a^{2}}\left(X^i_{t}\right)^{2} - 2 X^i_{t}\left(\frac{X^i_{t+1} + X^i_{t-1}}{a^{2}}\right) + \mu^2 {X^i_t}^2 -  \sum_{j\neq i}^{d} \left[X_{t}^{i}, X_{t}^{j}\right]^{2}\right],
\end{equation}

\begin{equation}
    \frac{\partial S_{\rm loc}}{\partial (X^i_t)_{a b}} = \frac{N a}{\lambda} \left\{ \frac{2}{a^2}(X^i_t)_{b a} - \left(\frac{X^i_{t+1} + X^i_{t-1}}{a^2}\right)_{b a} + \mu^2 \left({X^i_t}\right)_{b a} - \sum_{j\neq i}^d\left[X^j_t, \left[X^i_t, X^j_t\right]\right]_{b a} \right\}.
\end{equation}

\section{Simulation results}

\begin{figure}[hbt!]
\begin{center}
\includegraphics[scale=1.0]{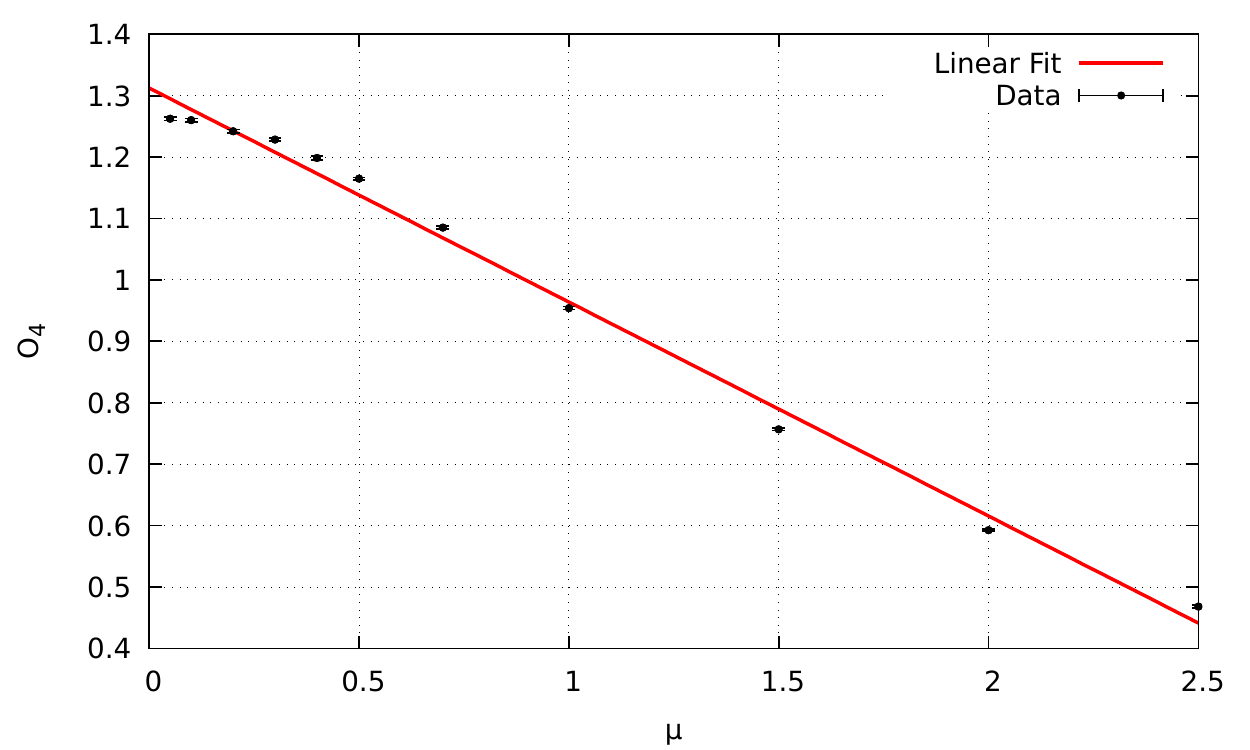}
\end{center}
\caption[Plot of average of $\left \langle O_4 \right \rangle$ against $\mu$.]{\label{fig:plot-O4}Plot of average of $\left \langle O_4 \right \rangle$ against $\mu$. Here, $d = 3$, $\lambda = 1$, $N = 8$, $T = 10$, $a = 0.5$ and $\beta = aT$. The linear extrapolation gives $\left. \left \langle O_4 \right \rangle \right|_{\mu = 0}= 1.31233 \pm 0.0113$.}
\label{fig:ch5-Ovsm}
\end{figure}

The model was simulated at $t = 0.2$ using the technique mentioned in the previous section. The data obtained are shown in Table. \ref{tab:data-O4} and the plot is provided in Fig. \ref{fig:plot-O4}.

\begin{table}[hbt!]
\caption[The values of observable $\left \langle O_4 \right \rangle$ for various $\mu$ values.]{\label{tab:data-O4}The average of $\left \langle O_4 \right \rangle$ at different $\mu$ values. Here, $d = 3$, $\lambda = 1$, $N = 8$, $T = 10$ and $a = 0.5$.}
\begin{center}
\begin{tabular}{ | c | c | } 
\hline 
$\mu$ & $\left \langle O_4 \right \rangle$ \\
  \hline \hline
0.05 &	1.26231	$\pm$ 0.00278846\\
0.1	&   1.26002	$\pm$ 0.00297276\\
0.2	&   1.24189	$\pm$ 0.00292772\\
0.3	&   1.22837	$\pm$ 0.00267511\\
0.4	&   1.19861	$\pm$ 0.00346254\\
0.5	&   1.16484	$\pm$ 0.00210862\\
0.7	&   1.08522	$\pm$ 0.0024943\\
1.0	&   0.954097 $\pm$ 0.00244354\\
1.5	&   0.756634 $\pm$ 0.00194958\\
2.0	&   0.592594 $\pm$ 0.00177286\\
2.5	&   0.468172 $\pm$ 0.00301384\\
  \hline
\end{tabular}
\end{center}
\end{table}

\clearpage

The value of $\left \langle O_4 \right \rangle$ at $\mu = 0$ obtained from a linear extrapolation is $1.31233 \pm 0.0113$. We also performed simulations at $\mu = 0$, fortunately, no flat directions were encountered. The value of $\left \langle O_4 \right \rangle$ came out to be $1.26909 \pm 0.00316178$ for $\mu = 0$. The difference is about $3.17\%$, which indicates that the linear fit is not a good choice to be used in the extrapolation and we should consider some other fit function.

\chapter{The $D = 4$ Model}
In this chapter, we look at the model after introducing a gauge field in the action. We call this model the $D = 4$ model. It has three `spatial dimensions' (encoded through the presence of the three scalar fields), a commutator potential and one gauge field. 

The Euclidean action for this model is given by
\begin{equation}
\label{eq:ch6-act} 
S_E = \frac{N}{\lambda} \int_{0}^{\beta} d t \operatorname{tr} \left\{\frac{1}{2}\left(\mathcal{D}_{t} X^{i}\right)^{2} - \sum_{i, j = 1}^{3} \frac{1}{4}\left[X^{i}, X^{j}\right]^{2}\right\}.
\end{equation}
Here
\begin{equation}
\mathcal{D}X^i = \frac{\partial X^i}{\partial t} - i[A(t), X^i] \nonumber
\end{equation}
is the covariant derivative and boundary conditions are periodic, $X(t + \beta) = X(t)$ and $A(t + \beta) = A(t)$. This model is similar to the bosonic BFSS model except that the BFSS model has nine spatial dimensions (appearing as nine scalar fields). The action in Eq. \eqref{eq:ch6-act} is invariant under the transformation 
\begin{equation}
\begin{array}{c}
X^{i}(t) \longrightarrow \Omega(t) X^{i}(t) \Omega^{\dagger}(t), \\
A(t) \longrightarrow \Omega(t)\left(A(t) + i \frac{d}{d t}\right) \Omega^{\dagger}(t),
\end{array}
\end{equation}
where $\Omega$ is a unitary matrix.

The partition function is given by
\begin{equation}
    Z = \int \mathcal{D}[A]\mathcal{D}[X] e^{-S_E}.
\end{equation}

As explained in chapter one, it is necessary to introduce link fields to put the action on the lattice. Using the discrete form of the covariant derivative as defined in Eq. \eqref{eq:ch1-covdr}, the lattice action is given by
\begin{equation}
S_{\rm lat} = \frac{N}{\lambda} \sum_{t=1}^T \operatorname{tr} \left\{-\frac{1}{a} X_{t}^{i} U_{t, t+1} X_{t+1}^{i} U_{t, t+1}^{\dagger} + \frac{1}{a}\left(X_{t}^{i}\right)^{2} - \sum_{i, j = 1}^3 \frac{a}{4} \left[X_{t}^{i}, X_{t}^{j}\right]^{2}\right\}.
\end{equation}

This action can be written in a much simpler form by using the $SU(N)$ local symmetry of the model. The details of the derivation are not included here and are given in Ref. \cite{Filev_2016}. The resulting reduced action is 
\begin{eqnarray}
S_{\rm lat}[X, D] &=& \frac{N}{\lambda} \operatorname{tr} \Bigg\{ -\frac{1}{a} \sum_{t=1}^{T-1} X_{t}^{i} X_{t+1}^{i} - \frac{1}{a} X_{T}^{i} D X_{1}^{i} D^{\dagger} \nonumber \\
&& ~~~~~~~~~~~~ + \sum_{t=1}^{T} \left[\frac{1}{a}\left(X_{t}^{i}\right)^{2} - \frac{a}{4}\left[X_{t}^{i}, X_{t}^{j}\right]^{2}\right] \Bigg\}.
\end{eqnarray}
Here $D = {\rm diag} (e^{i\theta_{1}}, \cdots, e^{i\theta_{N}})$ $\in SU(N)$ and $\theta_1, \cdots, \theta_N$ are the gauge variables which interacts only with the first and the last lattice site. The sums over $i\text{ and }j$ are implicit.

The partition function in the form of gauge variables is written as
\begin{equation}
Z \approx \int \prod_{k=1}^N d \theta_{k} \prod_{i=1}^3 \prod_{t=1}^T  d X_{t}^{i} e^{-(S_{\rm lat}\left[X, D(\theta)\right] + S_{\rm FP}[\theta])},
\end{equation}
where $S_{\rm FP}$ is the part of the action containing the Faddeev-Popov determinant
\begin{equation}
S_{\rm FP}[\theta] = -\sum_{l \neq m} \ln \left|\sin \left( \frac{\theta_{l} - \theta_{m}}{2} \right) \right|.
\end{equation}

\section{HMC for the model}

Since we have introduced $N$ number of new variables in the action, they also have to be updated during the molecular evolution. For fast convergence, it is better to update them together at every lattice site. The local action to work with according to this approach can be written as
\begin{equation}
    S^{\prime}_{\rm loc} = S_{\rm loc} + S_{\rm FP}[\theta],
\end{equation}
where 
\begin{equation}
S_{\rm loc} = \left \{
    \begin{array}{cl}
         & \bullet \text{ If $t = 1$}\\
         & \frac{N}{\lambda} \operatorname{tr}\left\{-\frac{1}{a} X_{1}^{i} X_{2}^{i}-\frac{1}{a} \sum_{i=1}^{3} X_{T}^{i} D X_{1}^{i} D^{\dagger}+\frac{1}{a}\left(X_{1}^{i}\right)^{2}-\frac{a}{2}\sum_{j\neq i}^{3}\left[X_{1}^{i}, X_{1}^{j}\right]^{2} \right\}, \\
         &\\
         & \bullet \text{ If $t = T$}\\
         & \frac{N}{\lambda} \operatorname{tr}\left\{-\frac{1}{a} X_{T-1}^{i} X_{T}^{i}-\frac{1}{a} \sum_{i=1}^{3} X_{T}^{i} D X_{1}^{i} D^{\dagger}+\frac{1}{a}\left(X_{T}^{i}\right)^{2}-\frac{a}{2}\sum_{j\neq i}^{3}\left[X_{T}^{i}, X_{T}^{j}\right]^{2} \right\}, \\
         &\\
         & \bullet \text{ If $t \neq 1$ and $ t \neq T$} \\
         &\frac{N}{\lambda} \operatorname{tr}\left\{-\frac{1}{a} \left(X_{t-1}^{i} X_{t}^{i} + X_{t}^{i} X_{t+1}^{i}\right)-\frac{1}{a} \sum_{i=1}^{3} X_{T}^{i} D X_{1}^{i} D^{\dagger}+\frac{1}{a}\left(X_{t}^{i}\right)^{2}\right.\\ &~~~~~~ \left.-\frac{a}{2}\sum_{j\neq i}^{3}\left[X_{t}^{i}, X_{t}^{j}\right]^{2} \right\}.
    \end{array}
    \right.
\end{equation}

The corresponding local Hamiltonian is 
\begin{equation}
    H = \frac{1}{2} \operatorname{tr}{P^{i}_{t}}^2 + \frac{1}{2}\sum_{l=1}^N P_{l} + S^{\prime}_{\rm loc}.
\end{equation}
Here $P^{i}_{t}$ is canonical momenta corresponding to the Hermitian matrix $X_t^i$ and $P_l$ is canonical momenta corresponding to the angles $\theta_l$.

The Hamilton's equation are 
\begin{equation}
\begin{aligned}
&(\dot{P^{i}_{t}} )_{l m} = -\partial S^{\prime}_{\rm loc} / \partial (X^{i}_{t})_{ m l}\quad, \quad \dot{P}_{l} = -\partial S^{\prime}_{\rm loc} / \partial \theta_{l}, \\
&(\dot{X^{i}_{t}})_{l m} = (P^{i}_{t})_{l m}, \quad \dot{\theta}_{l} = P_{l}.
\end{aligned}
\end{equation}

The following equations are used in the Leapfrog method
\begin{equation}
\begin{aligned}
- \partial S^{\prime}_{\rm loc} / \partial (X^i_{1})_{ m l} &= \frac{N}{\lambda a}\left(X_{2}^{i} - 2 X_{1}^{i} + D^{\dagger} X_{T}^{i} D\right)_{l m} + \frac{N a}{\lambda} \sum_{j=1}^{3}\left[X_{1}^{j}, \left[X_{1}^{i}, X_{1}^{j}\right]\right]_{l m}; \\
-\partial S^{\prime}_{\rm loc} / \partial (X^i_{t})_{ m l} &= \frac{N}{ \lambda a}\left(X_{t+1}^{i} - 2 X_{t}^{i} + X_{t-1}^{i}\right)_{l m} + \frac{N a}{\lambda}\sum_{j=1}^{3}\left[X_{t}^{j}, \left[X_{t}^{i}, X_{t}^{j}\right]\right]_{l m} \\
& \text{ for } \quad t = 2, \ldots, T-1; \\
-\partial S^{\prime}_{\rm loc} / \partial (X_{T})_{m l} &= \frac{N}{\lambda a}\left(D X_{1}^{i} D^{\dagger}-2 X_{T}^{i}+X_{T-1}^{i}\right)_{l m}+\frac{N a}{\lambda}\sum_{j=1}^{3}\left[X_{T}^{j},\left[X_{T}^{i}, X_{T}^{j}\right]\right]_{l m}; \\
-\partial S^{\prime}_{\rm loc} / \partial \theta_{l} &= \frac{2 N}{\lambda a} \sum_{m=1}^{N} \Re \left[i (X^{i}_{T})_{ml} (X_{1}^{i})_{lm} e^{i\left(\theta_{l}-\theta_{m}\right)}\right]+\sum_{m, m \neq l} \cot \left(\frac{\theta_{l}-\theta_{m}}{2}\right),
\end{aligned}
\end{equation}
where $\Re$ indicates the real part.

\subsection{\texorpdfstring{Constraint for $\theta_{l}$}{}}

The link field $D$ is an element of the $SU(N)$ group, so the gauge variables $\theta_{l}$ follow the constraint 
\begin{equation}
    \sum_{k=1}^{N} \theta_{k}=0.
\end{equation}

\section{Observables}

The observables namely, the Polyakov loop, the extent of space, and the internal energy have been calculated to study the phase structure of the system. These quantities have the following definitions. \\

$\bullet$ Polyakov loop

\begin{equation}
            P = \frac{1}{N} \operatorname{tr} \mathcal{P} \left(e^{i \oint A}\right) = \frac{1}{N}\left| \sum_{k=1}^N e^{i\theta_{k}}\right|.
\end{equation}

$\bullet$ Extent of space

\begin{equation}
        \left \langle R^2 \right \rangle = \left\langle\frac{1}{N \beta} \int_{0}^{\beta} d t \operatorname{Tr}\left(X^{i}\right)^{2}\right\rangle.
\end{equation}

$\bullet$ Internal energy

\begin{equation}
        \frac{E}{N^2} = \left \langle -\frac{3}{4 N \beta} \int_{0}^{\beta} d t \operatorname{Tr} \left(\left[X^{i}, X^{j}\right]^{2}\right) \right \rangle.
\end{equation}

Note: These definitions stand for the case when $\lambda = 1$.

\section{Save and restart strategy}

The code to simulate the model can take a long time to complete the run, and if somehow the simulations break in between, for instance, when a power failure is encountered, then the code has to run from the beginning. It means one has to wait again for the system to thermalize. Therefore, to avoid such a problem, the code has to be designed in a manner that it saves the final configuration in a file which can later be read to continue the simulations directly from that point onwards.  For example, a code can be designed such that it saves the final configuration at the end of every 100 sweeps and then it would read in the saved configuration file to continue the run for the next 100 sweeps. In this way, the code can be made to run for a longer time to generate enough statistics on the data.

\section{Simulation results}

The code was set to run for different temperature values, each with a total of 1.6 million sweeps. The results obtained are given below. Our simulation results are in good agreement with the results given in Ref. \cite{Hanada_2007}.

\subsection{Polyakov loop}

\begin{figure}[hbt!]
\begin{center}
\includegraphics[scale=1.0]{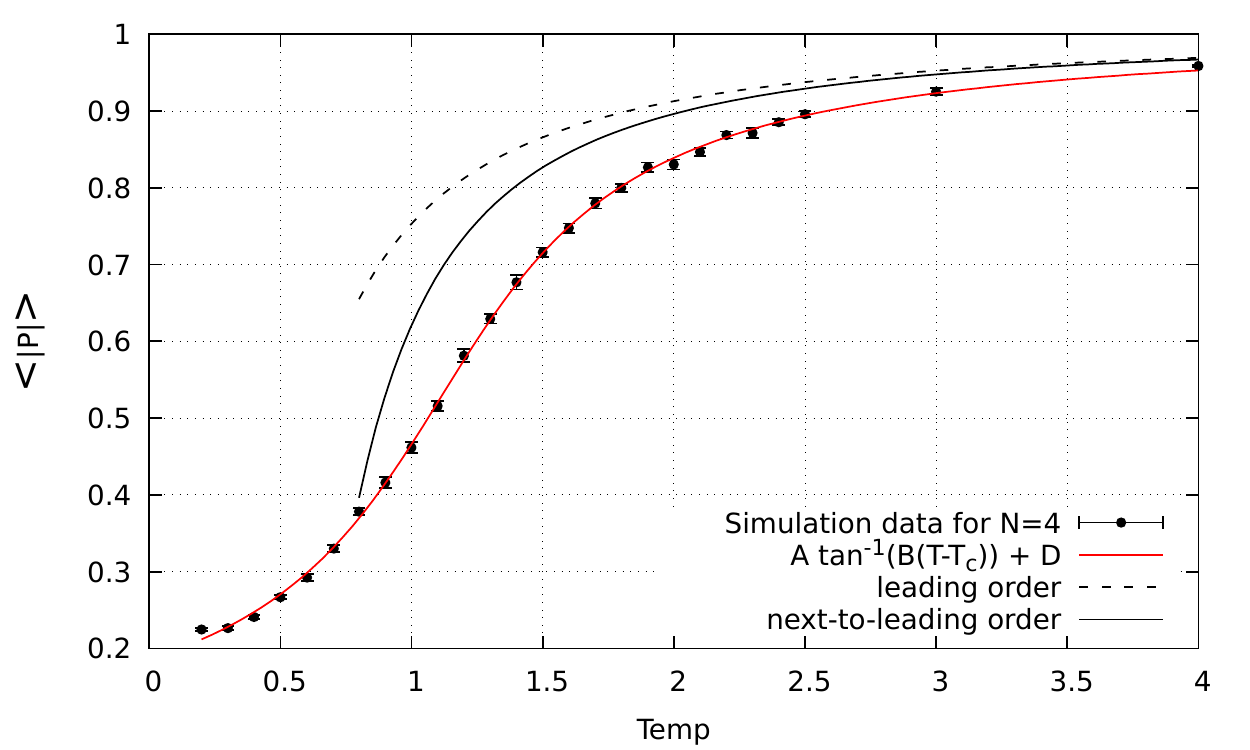}
\end{center}
\caption[The Polyakov loop observable for the $D = 4$ model.]{A plot of the expectation value of the Polyakov loop as a function of the temperature. Here, $d = 3$, $\lambda = 1$, $N = 4$ and $T = 20$. The data indicate a phase transition near $t \approx 1.11$.}
\label{fig:ch6-polyakov}
\end{figure}

The expectation value of the Polyakov loop, as discussed in Chapter 1, plays the role of an order parameter for the confining-deconfining phase transition. Fig. \ref{fig:ch6-polyakov} displays the expectation value of this parameter as a function of temperature. The change in slope of the curve near $t \approx 1.11$ shows the existence of a phase transition. Analytical results for high-temperature behavior of this quantity is given in Ref. \cite{Kawahara_2007}. The plot shows that the fitted curve to the data meets with the analytical results around $t = 4$. The data is fitted with the function 
\begin{equation}
\label{eq:ch6-fit}
A \tan^{-1} (\,B \,(T-T_c)\,) + D, 
\end{equation} 
and the obtained values of parameters $A, B, T_c$ and $D$ are given in Table. \ref{tb:ch6-1}.

\begin{table}[hbt!]
\caption[The values of the fit parameters for the Polyakov loop in the $D = 4$ model.]{The values of the fit parameters $A, B, T_c$ and $D$. Here, $d = 3$, $\lambda = 1$, $N = 4$ and $T = 20$.}
\label{tb:ch6-1}
\begin{center}
\begin{tabular}{ | c | c | } 
\hline
Parameter & Value \\
  \hline \hline
    A	& 0.3083 $\pm$ 0.0041 \\
    B	& 1.7967 $\pm$ 0.0453 \\
    $T_{c}$	& 1.1125 $\pm$ 0.0083 \\
    D   & 0.5272 $\pm$ 0.0029 \\
  \hline
\end{tabular}
\end{center}
\end{table}

\begin{figure}[hbt!]
\begin{center}
\includegraphics[scale=0.9]{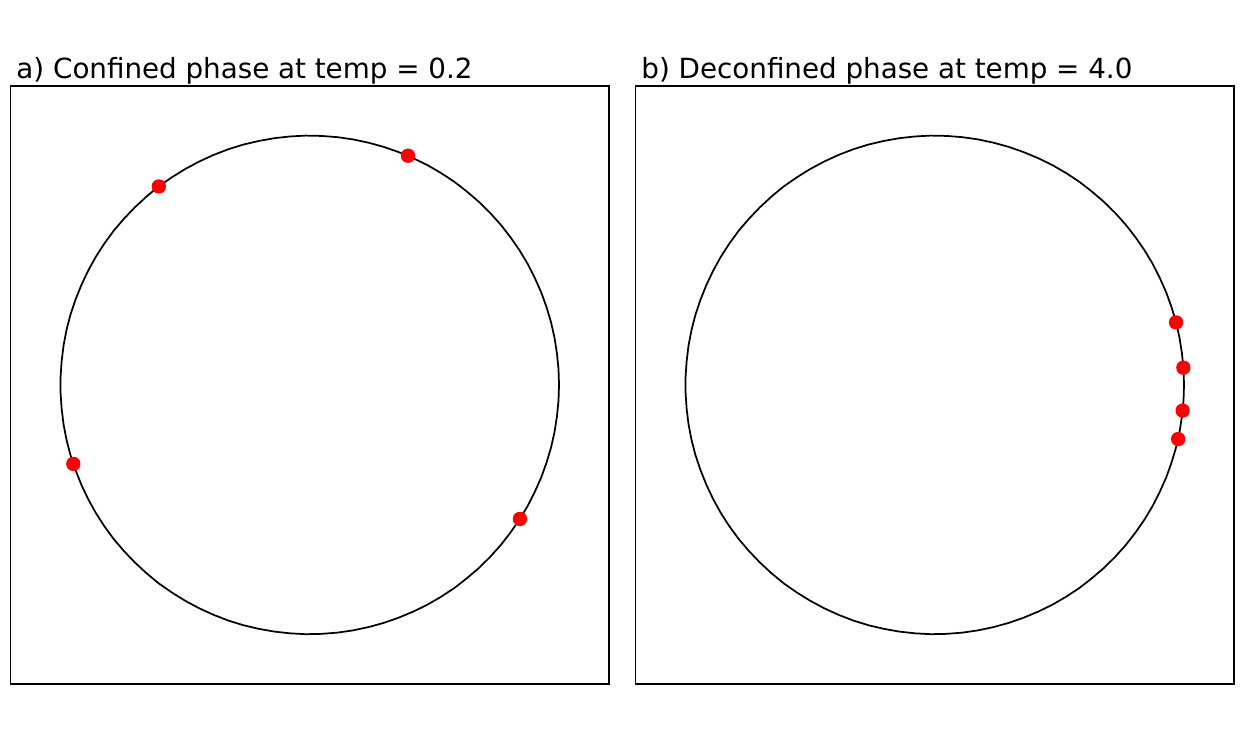}
\end{center}
\caption[The eigenvalue distribution for the $D = 4$ model.]{The distribution of eigenvalues of the Polyakov loop operator, on a unit circle in the complex plane, for the $D = 4$ model. The simulations are performed for $N = 4$.}
\label{fig:ch6-eigenvalue}
\end{figure}

The distributions of the eigenvalues of the Polyakov loop operator in the two phases (confined and deconfined) of the model are shown in Fig. \ref{fig:ch6-eigenvalue}. The eigenvalues spread uniformly on the unit circle in the confined phase, whereas they cluster about one point in the deconfined phase.

\subsection{Internal energy and extent of space}

\begin{figure}[hbt!]
\begin{center}
\includegraphics[scale=1.0]{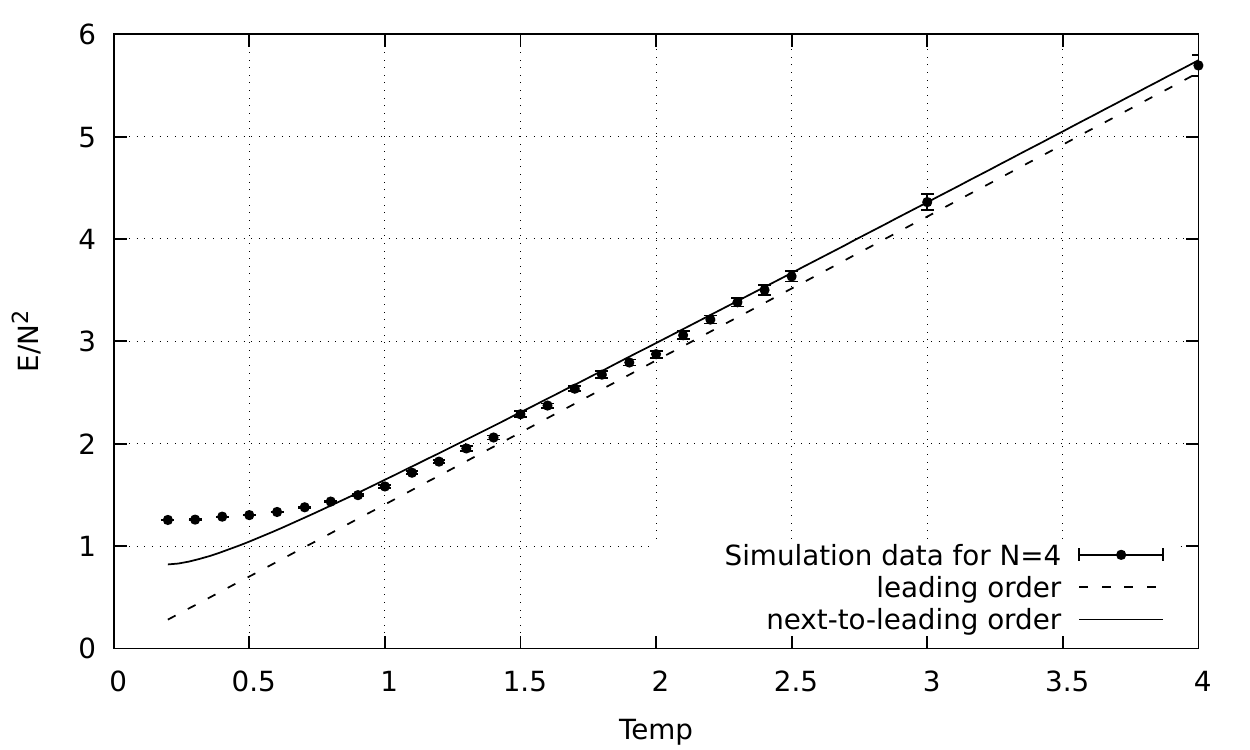}
\end{center}
\caption[Internal energy for the $D = 4$ model.]{Plot of the scaled energy as a function of the temperature. Here, $d = 3$, $\lambda = 1$, $N = 4$ and $T = 20$. The data suggest the existence of a phase transition near $t = 1.1$.}
\label{fig:ch6-energy}
\end{figure}

\begin{figure}[hbt!]
\begin{center}
\includegraphics[scale=1.0]{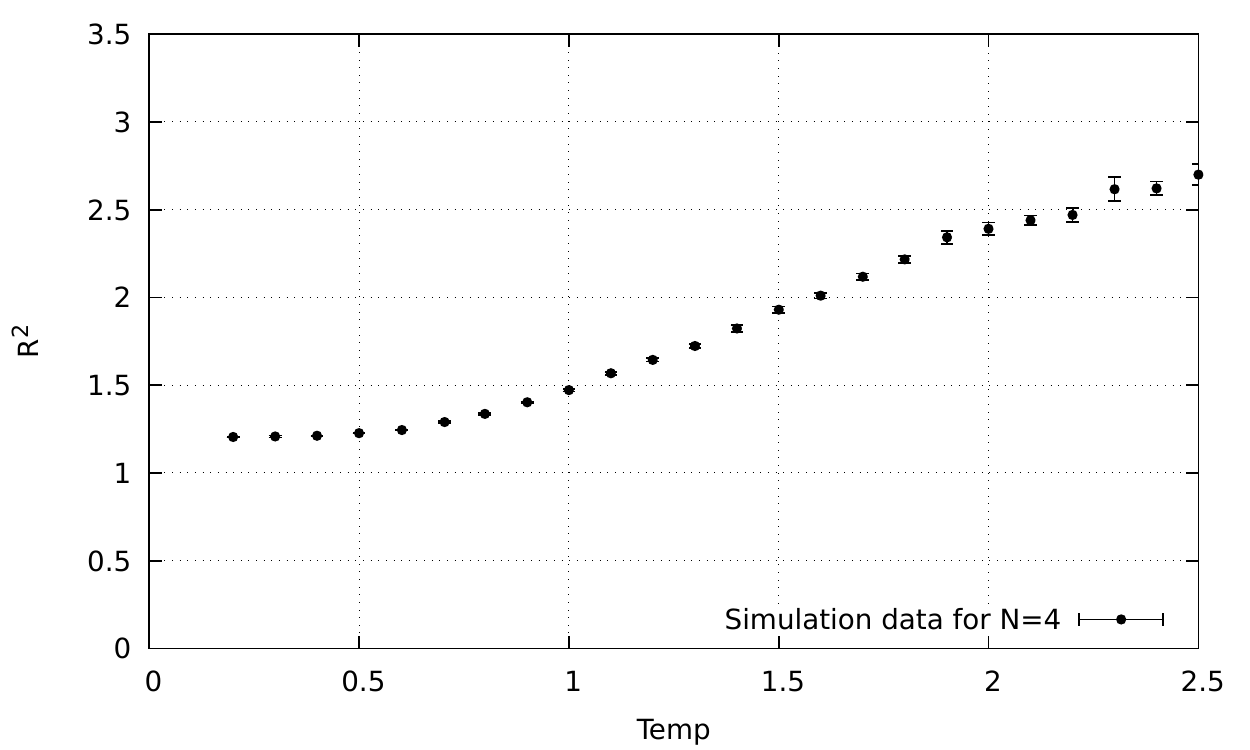}
\end{center}
\caption[Extent of space for the $D = 4$ model.]{Plot of the extent of space observable as a function of the temperature. Here, $d = 3$, $\lambda = 1$, $N = 4$ and $T = 20$.}
\label{fig:ch6-R2}
\end{figure}

In Figs. \ref{fig:ch6-energy} and \ref{fig:ch6-R2} we show the plots for scaled internal energy $\left \langle E/N^2 \right \rangle$ and the ``extent of space'' $\left \langle R^2 \right \rangle$ as functions of temperature. It can be seen that the data overlaps with the analytical results for the high-temperature behavior for $t \gtrsim 2$.

\clearpage

We expect similar behavior for these quantities in the quenched BFSS matrix model also.

\chapter{Bosonic BFSS Matrix Model}
In the previous chapter we set up a framework to include the gauge field with the commutator potential. We can now move on to simulate the quenched BFSS model. The Euclidean action of the model is 
\begin{equation} 
S_{\rm E}=\frac{N}{\lambda} \int_{0}^{\beta} d t \operatorname{tr}\left\{\frac{1}{2}\left(\mathcal{D}_{t} X^{i}\right)^{2}-\sum_{i,j=1}^{9}\frac{1}{4}\left[X^{i}, X^{j}\right]^{2}\right\}.
\end{equation}

Here
\begin{equation}
\mathcal{D}X^i = \frac{\partial X^i}{\partial t} - i[A(t),X^i] \nonumber
\end{equation}
is the covariant derivative and boundary conditions are periodic: $X(t+\beta) = X(t)$ and $A(t+\beta) = A(t)$. \\

As mentioned in the previous chapter, the action can be written in a simpler form. The simplified action is 
\begin{eqnarray}
S_{\rm lat}[X, D] &=& \frac{N}{\lambda} \operatorname{tr} \Bigg\{ - \frac{1}{a} \sum_{t=1}^{T-1} X_t^i X_{t+1}^i - \frac{1}{a} X_{T}^{i} D X_{1}^{i} D^{\dagger} \nonumber \\
&& ~~~~~~~~ + \sum_{t=1}^T \left[\frac{1}{a} \left( X_{t}^i \right)^2 - \frac{a}{4} \left[X_t^i, X_t^j \right]^2 \right] \Bigg\},
\end{eqnarray}
where $D = {\rm diag} (e^{i\theta_{1}}, \cdots, e^{i\theta_{N}})$ $\in SU(N)$. The sums over indices $i$ and $j$ are implicit and they run from $1$ to $9$.

\section{HMC for the model}

The local action is
\begin{equation}
    S^{\prime}_{\rm loc} = S_{\rm loc} + S_{\rm FP}[\theta] ,
\end{equation}
where,
\begin{equation}
\label{eq:ch7-locaction}
S_{\rm loc} = \left \{
    \begin{array}{cl}
         & \bullet \text{ If $t = 1$}\\
         & \frac{N}{\lambda} \operatorname{tr}\left\{-\frac{1}{a} X_{1}^{i} X_{2}^{i}-\frac{1}{a} \sum_{i=1}^{9} X_{T}^{i} D X_{1}^{i} D^{\dagger}+\frac{1}{a}\left(X_{1}^{i}\right)^{2}-\frac{a}{2}\sum_{j\neq i}^{9}\left[X_{1}^{i}, X_{1}^{j}\right]^{2}\right\},\\
         &\\
         & \bullet \text{ If $t = T$}\\
         & \frac{N}{\lambda} \operatorname{tr}\left\{-\frac{1}{a} X_{T-1}^{i} X_{T}^{i}-\frac{1}{a} \sum_{i=1}^{9} X_{T}^{i} D X_{1}^{i} D^{\dagger}+\frac{1}{a}\left(X_{T}^{i}\right)^{2}-\frac{a}{2}\sum_{j\neq i}^{9}\left[X_{T}^{i}, X_{T}^{j}\right]^{2}\right\},\\
         &\\
         & \bullet \text{ If $t \neq$ 1 and $t \neq T$}\\
         &\frac{N}{\lambda} \operatorname{tr}\left\{-\frac{1}{a} \left(X_{t-1}^{i} X_{t}^{i} + X_{t}^{i} X_{t+1}^{i}\right)-\frac{1}{a} \sum_{i=1}^{9} X_{T}^{i} D X_{1}^{i} D^{\dagger}+\frac{1}{a}\left(X_{t}^{i}\right)^{2}\right.\\ &~~~~~~ \left.-\frac{a}{2}\sum_{j\neq i}^{9}\left[X_{t}^{i}, X_{t}^{j}\right]^{2}\right\}
    \end{array}
    \right.
\end{equation}

The equations of motion are 
\begin{equation}
\begin{aligned}
&(\dot{P^{i}_{t}} )_{l m}=-\partial S^{\prime}_{\rm loc} / \partial (X^{i}_{t})_{ m l}\quad, \quad \dot{P}_{l}=-\partial S^{\prime}_{\rm loc} / \partial \theta_{l},\\
&(\dot{X^{i}_{t}})_{l m}=(P^{i}_{t})_{l m}\quad , \quad \dot{\theta}_{l}=P_{l}.
\end{aligned}
\end{equation}

In the Leapfrog algorithm, the following equations are used.
\begin{equation}
\label{eq:ch7-leapfrog}
\begin{aligned}
-\partial S^{\prime}_{\rm loc} / \partial (X^i_{1})_{ m l} &= \frac{N}{\lambda a}\left(X_{2}^{i} - 2 X_{1}^{i} + D^{\dagger} X_{T}^{i} D\right)_{l m} + \frac{N a}{\lambda}\sum_{j=1}^{9}\left[X_{1}^{j},\left[X_{1}^{i}, X_{1}^{j}\right]\right]_{l m}; \\
-\partial S^{\prime}_{\rm loc} / \partial (X^i_{t})_{ m l} &= \frac{N}{ \lambda a}\left(X_{t+1}^{i} - 2 X_{t}^{i}+X_{t-1}^{i}\right)_{l m} + \frac{N a}{\lambda}\sum_{j=1}^{9}\left[X_{t}^{j},\left[X_{t}^{i}, X_{t}^{j}\right]\right]_{l m}\\
& \text{ for } \quad t = 2, \ldots, T-1; \\
-\partial S^{\prime}_{\rm loc} / \partial (X_{T})_{m l} &= \frac{N}{\lambda a}\left(D X_{1}^{i} D^{\dagger} - 2 X_{T}^{i}+X_{T-1}^{i}\right)_{l m} + \frac{N a}{\lambda}\sum_{j=1}^{9}\left[X_{T}^{j},\left[X_{T}^{i}, X_{T}^{j}\right]\right]_{l m}; \\
-\partial S^{\prime}_{\rm loc} / \partial \theta_{l} &= \frac{2 N}{\lambda a} \sum_{m=1}^{N} \Re\left[i (X^{i}_{T})_{ml} (X_{1}^{i})_{lm} e^{i\left(\theta_{l}-\theta_{m}\right)}\right]+\sum_{m, m \neq l} \cot \left(\frac{\theta_{l}-\theta_{m}}{2}\right).
\end{aligned}
\end{equation}

\section{Simulation results}

\begin{figure}[hbt!]
\begin{center}
\includegraphics[scale=1.0]{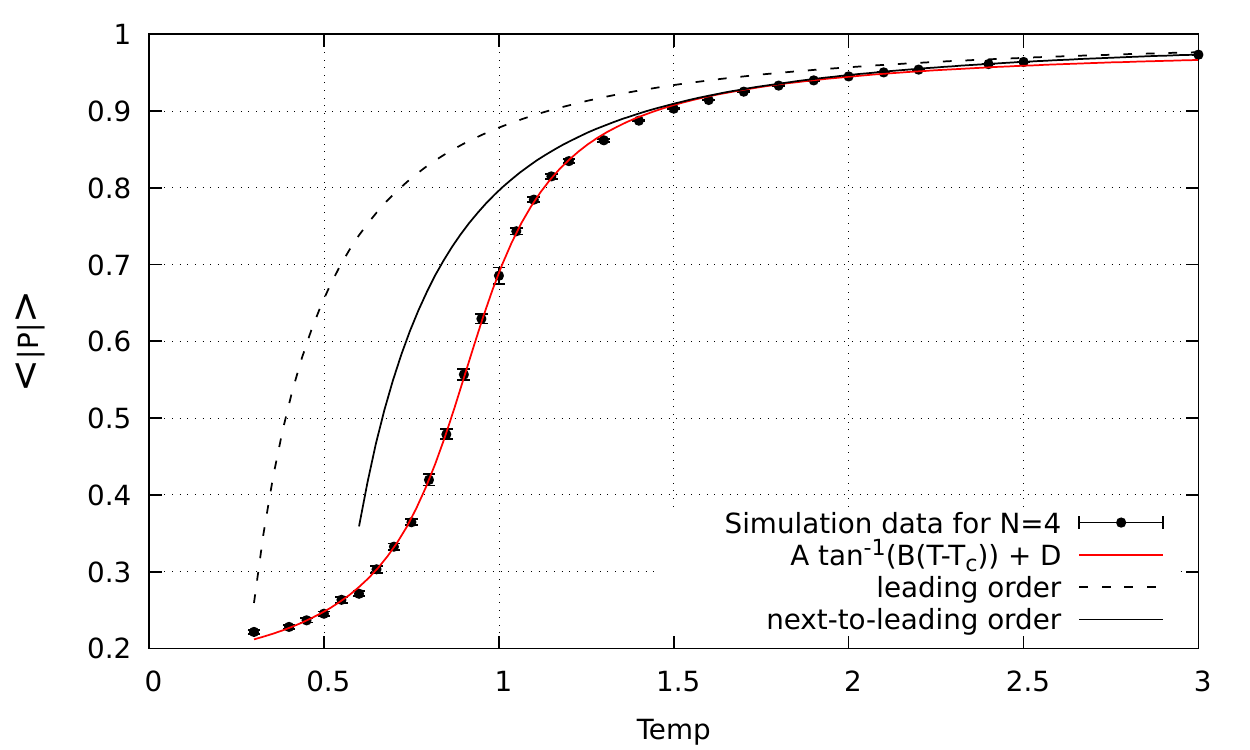}
\end{center}
\caption[Polyakov loop for the bosonic BFSS model.]{A plot of the expectation value of the Polyakov loop as a function of the temperature. Here, $d = 9$, $\lambda = 1$, $N = 4$ and $T = 10$. The plot displays a phase transition near $t \approx 0.90$. The broken line and the solid line denote the results of the high-temperature expansion for $N = 4$, which are obtained in Ref. \cite{Kawahara_2007} at the leading order and at the next-to-leading order, respectively.}
\label{fig:ch7-polyakov}
\end{figure}

\begin{figure}[hbt!]
\begin{center}
\includegraphics[scale=1.0]{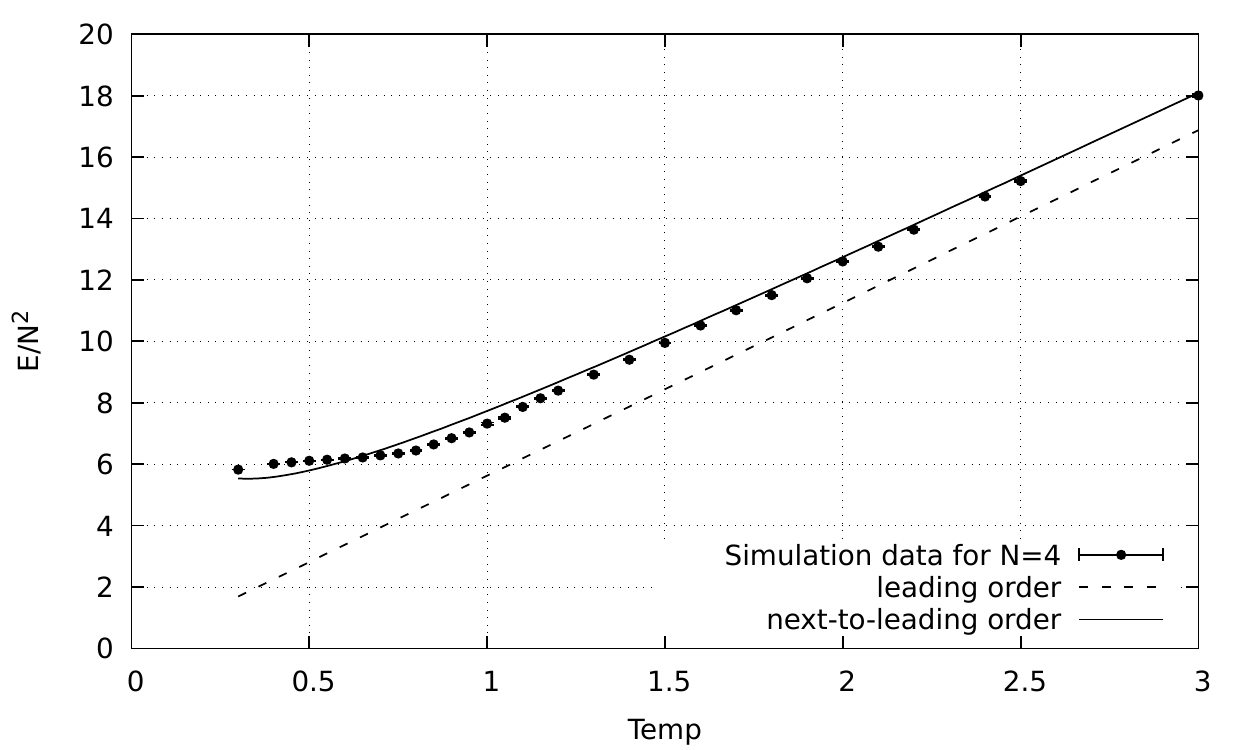}
\end{center}
\caption[Internal energy for the bosonic BFSS model.]{Figure displays a plot of the scaled energy as a function of the temperature for $N = 4$ and $T = 10$. Here, $d = 9$ and $\lambda = 1$. The plot suggests the existence of a phase transition near $t \approx 0.90$.}
\label{fig:ch7-energy}
\end{figure}

\begin{figure}[hbt!]
\begin{center}
\includegraphics[scale=1.0]{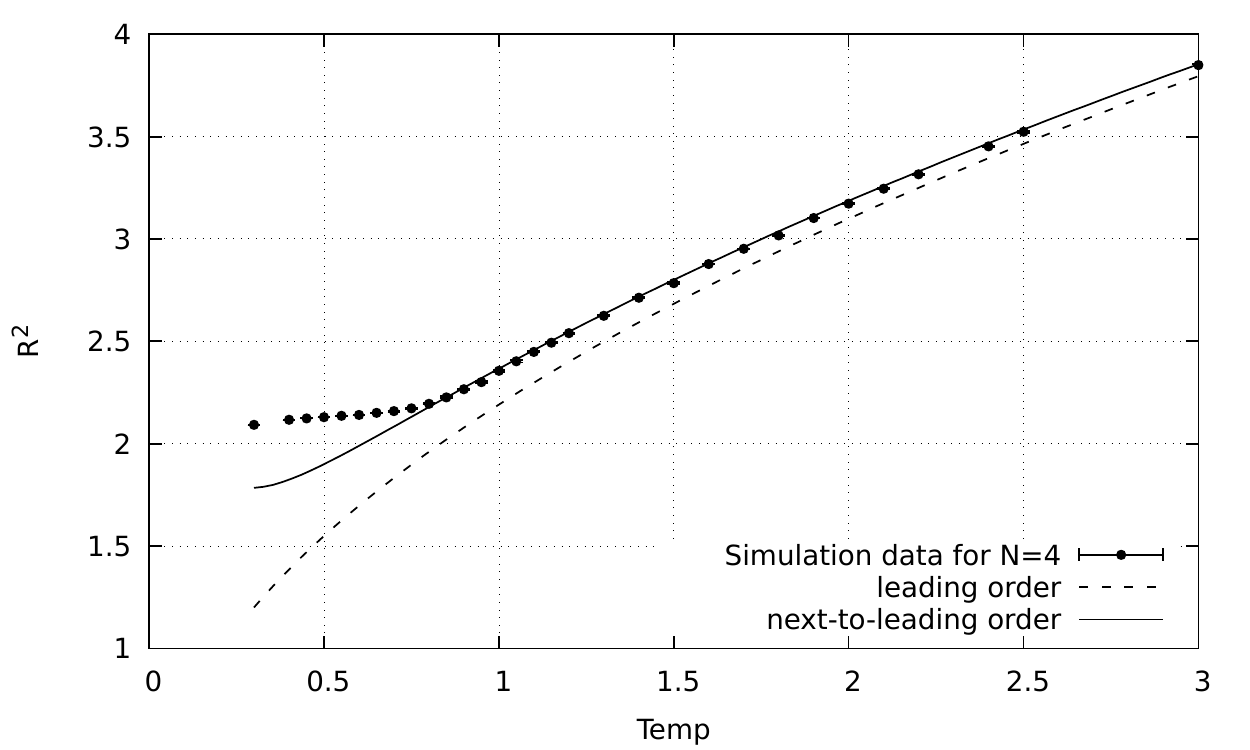}
\end{center}
\caption[Extent of space for the bosonic BFSS model.]{The extent of space is plotted against the temperature. Here, $d = 9$, $\lambda = 1$, $N = 4$ and $T = 10$.}
\label{fig:ch7-R2}
\end{figure}

Simulation results given in Figs. \ref{fig:ch7-polyakov}, \ref{fig:ch7-energy} and \ref{fig:ch7-R2} indicate that a phase transition exists around $t \approx 0.90$. The value of the critical temperature obtained in our simulations is in agreement with those obtained in Refs. \cite{Kawahara_2007_Oct} and \cite{Filev_2016}.

\clearpage

\subsection{Polyakov loop}

The values of the parameters $A$, $B$, $T_C$ and $D$ are given in Table. \ref{tb:ch7-1}.

\begin{table}[hbt!]
\caption[Fit parameters for the Polyakov loop in the bosonic BFSS model.]{The values of the fit parameters $A, B, T_c$ and $D$. Here, $d = 9$, $\lambda = 1$, $N = 4$ and $T = 10$.}
\label{tb:ch7-1}
\begin{center}
\begin{tabular}{ | c | c | } 
\hline
Parameter & Value \\
  \hline \hline
    A	& 0.273982 $\pm$ 0.001569\\
    B	& 5.37305 $\pm$ 0.08925\\
    $T_{c}$	& 0.905404 $\pm$ 0.00201\\
    D   & 0.560378 $\pm$ 0.001439\\
  \hline
\end{tabular}
\end{center}
\end{table}

\clearpage

\subsection{Internal energy and extent of space}

\chapter{Bosonic BMN Matrix Model}
The BMN matrix model can be considered as an extension of the BFSS matrix model: it has additional mass terms besides the derivative and the commutator terms.
 
The Euclidean action of the matrix model without the fermions is
\begin{equation}
\begin{aligned}
S_{\rm E} = & \,\frac{N}{\lambda} \int_{0}^{\beta} d t \operatorname{tr} \left[\frac{1}{2}\left(D_{t} X^{i}\right)^{2}-\frac{1}{4}\left[X^{i}, X^{j}\right]^{2} + \frac{1}{2}\left(\frac{\mu}{3}\right)^{2}\left(X^{I}\right)^{2} + \frac{1}{2}\left(\frac{\mu}{6}\right)^{2}\left(X^{I^{\prime}}\right)^{2}\right. \\ & \left. ~~~~~~ + \, i \frac{\mu}{3} \epsilon_{I J K} X^{I} X^{J} X^{K}\right] ,
\end{aligned}
\end{equation}
where $i, j = 1, \cdots, 9$, $I^{\prime} = 4, \cdots, 9$ and $I, J, K = 1, 2, 3$. \\ 

The last term in the action can also be written in a simplified form
\begin{equation}
    i \frac{\mu}{3} \epsilon_{I J K} \operatorname{tr}\left(X^{I} X^{J} X^{K}\right) = i \mu \operatorname{tr}\left(X^{1} .\left[X^{2}, X^{3}\right]\right).
\end{equation}

\section{HMC for the model}

We can use the same approach employed in the BFSS model here as well. The extra task is just to add the mass terms and their gradients in Eqs. \eqref{eq:ch7-locaction} and \eqref{eq:ch7-leapfrog}, respectively. \\

The local action along with the mass terms is given by

\begin{equation}
    S^{\prime}_{\rm loc} = S_{\rm loc} + S_{\rm FP}[\theta],
\end{equation}

where 

\begin{equation}
S_{\rm loc} = \left \{
    \begin{array}{cl}
         & \bullet \text{ If $t = 1$}\\
         & \frac{N}{\lambda} \operatorname{tr}\left\{-\frac{1}{a} X_{1}^{i} X_{2}^{i}-\frac{1}{a} \sum_{i=1}^{9} X_{T}^{i} D X_{1}^{i} D^{\dagger}+\frac{1}{a}\left(X_{1}^{i}\right)^{2}-\frac{a}{2}\sum_{j\neq i}^{9}\left[X_{1}^{i}, X_{1}^{j}\right]^{2} \right\} + {\rm M} , \\
         &\\
         & \bullet \text{ If $t = T$}\\
         & \frac{N}{\lambda} \operatorname{tr}\left\{-\frac{1}{a} X_{T-1}^{i} X_{T}^{i}-\frac{1}{a} \sum_{i=1}^{9} X_{T}^{i} D X_{1}^{i} D^{\dagger}+\frac{1}{a}\left(X_{T}^{i}\right)^{2}-\frac{a}{2}\sum_{j\neq i}^{9}\left[X_{T}^{i}, X_{T}^{j}\right]^{2} \right\} + {\rm M} , \\
         &\\
         & \bullet \text{ If $t \neq 1$ and $ t \neq T$} \\
         &\frac{N}{\lambda} \operatorname{tr}\left\{-\frac{1}{a} \left(X_{t-1}^{i} X_{t}^{i} + X_{t}^{i} X_{t+1}^{i}\right)-\frac{1}{a} \sum_{i=1}^{9} X_{T}^{i} D X_{1}^{i} D^{\dagger}+\frac{1}{a}\left(X_{t}^{i}\right)^{2}\right.\\ &~~~~~~ \left.-\frac{a}{2}\sum_{j\neq i}^{9}\left[X_{t}^{i}, X_{t}^{j}\right]^{2} \right\} + {\rm M} ,
    \end{array}
    \right.
\end{equation}

with

\begin{equation}
    {\rm M} = \left \{
    \begin{array}{cl}
         &  \frac{N}{\lambda} \operatorname{tr}\left\{\frac{a}{2}\left(\frac{\mu}{3}\right)^{2} \left(X^{i}_{t}\right)^{2} +  i a \mu \left(X^{1}_{t} .\left[X^{2}_{t}, X^{3}_{t}\right]\right) \right\} \quad \text{for $i=1,2,3$;}\\
         &  \frac{N}{\lambda} \operatorname{tr}\left\{\frac{a}{2}\left(\frac{\mu}{6}\right)^{2} \left(X^{i}_{t}\right)^{2} \right\} \quad  \text{for $i=4, \cdots, 9$}.
    \end{array}
    \right.
\end{equation}
\\
The Faddeev-Popov term is the same as that of the BFSS matrix model. \\

The equations for negated gradient of $S^{\prime}_{\rm loc}$ are the following

\begin{equation}
\begin{aligned}
-\partial S^{\prime}_{\rm loc} / \partial (X^i_{1})_{ m l} &= \frac{N}{\lambda a}\left(X_{2}^{i} - 2 X_{1}^{i} + D^{\dagger} X_{T}^{i} D\right)_{l m} + \frac{N a}{\lambda}\sum_{j=1}^{9}\left[X_{1}^{j},\left[X_{1}^{i}, X_{1}^{j}\right]\right]_{l m} + ({\rm M}^{\prime})_{l m}; \\
-\partial S^{\prime}_{\rm loc} / \partial (X^i_{t})_{ m l} &= \frac{N}{ \lambda a}\left(X_{t+1}^{i} - 2 X_{t}^{i}+X_{t-1}^{i}\right)_{l m} + \frac{N a}{\lambda}\sum_{j=1}^{9}\left[X_{t}^{j},\left[X_{t}^{i}, X_{t}^{j}\right]\right]_{l m} + ({\rm M}^{\prime})_{l m} \\
& \text{ for } \quad t = 2, \ldots, T-1; \\
-\partial S^{\prime}_{\rm loc} / \partial (X_{T})_{m l} &= \frac{N}{\lambda a}\left(D X_{1}^{i} D^{\dagger} - 2 X_{T}^{i}+X_{T-1}^{i}\right)_{l m} + \frac{N a}{\lambda}\sum_{j=1}^{9}\left[X_{T}^{j},\left[X_{T}^{i}, X_{T}^{j}\right]\right]_{l m} + ({\rm M}^{\prime})_{l m}; \\
-\partial S^{\prime}_{\rm loc} / \partial \theta_{l} &= \frac{2 N}{\lambda a} \sum_{m=1}^{N} \Re\left[i (X^{i}_{T})_{ml} (X_{1}^{i})_{lm} e^{i\left(\theta_{l}-\theta_{m}\right)}\right]+\sum_{m, m \neq l} \cot \left(\frac{\theta_{l}-\theta_{m}}{2}\right);
\end{aligned}
\end{equation}

where 

\begin{equation}
({\rm M}^{\prime})_{l m} = \frac{\partial {\rm M}}{\partial (X^{i}_{t})_{ m l } } = \left \{
    \begin{array}{cl}
         &  - \frac{N a }{\lambda} (\frac{\mu}{3})^{2} \left(X^{i}_{t}\right)_{l m} - \left \{\begin{array}{cl}
              &  \frac{i N  a \mu}{\lambda} \left[X^{2}_{t},X^{3}_{t}\right]_{l m} \quad \text{for $i=1$;}\\
              &  \frac{i N  a \mu}{\lambda} \left[X^{3}_{t},X^{1}_{t}\right]_{l m} \quad \text{for $i=2$;}\\
              &  \frac{i N a \mu}{\lambda} \left[X^{1}_{t},X^{2}_{t}\right]_{l m} \quad \text{for $i=3$;}
         \end{array} \right. \\
         &  - \frac{N a }{\lambda} (\frac{\mu}{6})^{2} \left(X^{i}_{t}\right)_{l m} \quad \text{for $i=4, \cdots, 9$}.
    \end{array}
    \right.
\end{equation}

\section{Observables}

One additional observable has also been calculated. It is defined below. \\

$\bullet$ Myers term

\begin{equation}
    {\rm Myr} = \left\langle\frac{i}{3 N \beta} \int_{0}^{\beta} d t \,\epsilon_{I J K} \operatorname{Tr}\left(X^{I} X^{J} X^{K}\right)\right\rangle.
\end{equation}

With the mass terms included, the internal energy is given by

$\bullet$ Internal energy
\begin{equation}
\begin{aligned}
        \frac{E}{N^2} & =  \left \langle \frac{1}{N \beta} \int_{0}^{\beta} d t \operatorname{Tr} \left[ -\frac{3}{4}\left[X^{i}, X^{j}\right]^{2}  + \left(\frac{\mu}{3}\right)^{2} \left(X^{I}\right)^{2} \right. \right. \\ & \left. \left. ~~~~~~~~ + \left(\frac{\mu}{6}\right)^{2} \left(X^{I^{\prime}}\right)^{2} +\, i \frac{5\mu}{6} \epsilon_{I J K} X^{I} X^{J} X^{K} \right] \right \rangle.
\end{aligned}
\end{equation}

Note : $\lambda = 1$ for above mentioned definitions.

\section{Simulation results}

\subsection{Symmetry breaking}

\begin{figure}[hbt!]
\begin{center}
\includegraphics[scale=1.0]{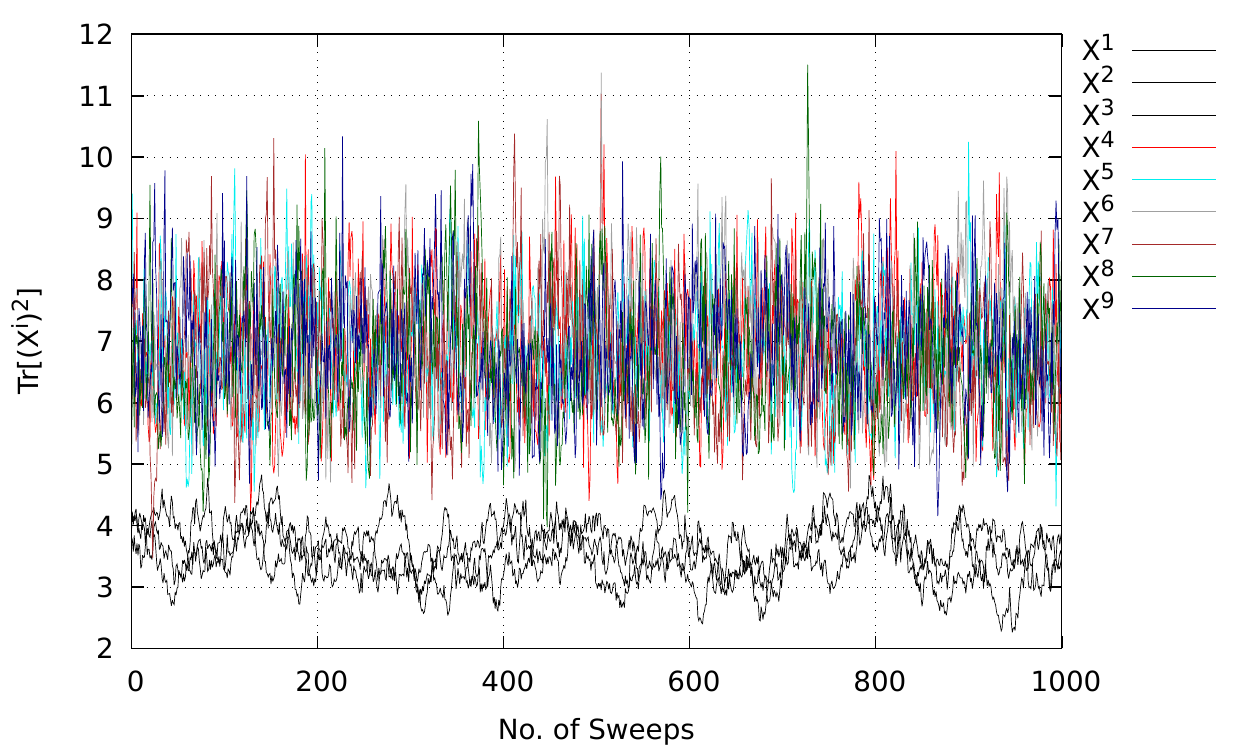}
\end{center}
\caption[$SO(9)$ symmetry breaking in the bosonic BMN model.]{A plot of $\operatorname{Tr}[{X^{i}}^{2}]$, $i = 1, 2, \cdots, 9$, against the number of Monte Carlo sweeps. It displays $SO(9)\xrightarrow{} SO(6)\times SO(3)$ global symmetry breaking in the BMN matrix model (six colored lines above and three black lines below). Here, $d = 9$, $\lambda = 1$, $N = 4$, $T = 10$, $t = 0.3$ and $\mu = 12$.}
\label{fig:ch8-symmetry}
\end{figure}

The six colored and three black lines at the top and bottom respectively in Fig. \ref{fig:ch8-symmetry} illustrate the expected $SO(9)\xrightarrow{} SO(6)\times SO(3)$ global symmetry breaking.

\clearpage

\subsection{\texorpdfstring{Observables for $\mu = 2.0$}{} }

The figures shown in this section demonstrate that the system undergoes a phase transition near $t \approx 0.91$. The same value was obtained for the critical temperature in Ref. \cite{Asano:2020yry}.

\subsubsection{Polyakov loop}

\begin{figure}[hbt!]
\begin{center}
\includegraphics[scale=1.0]{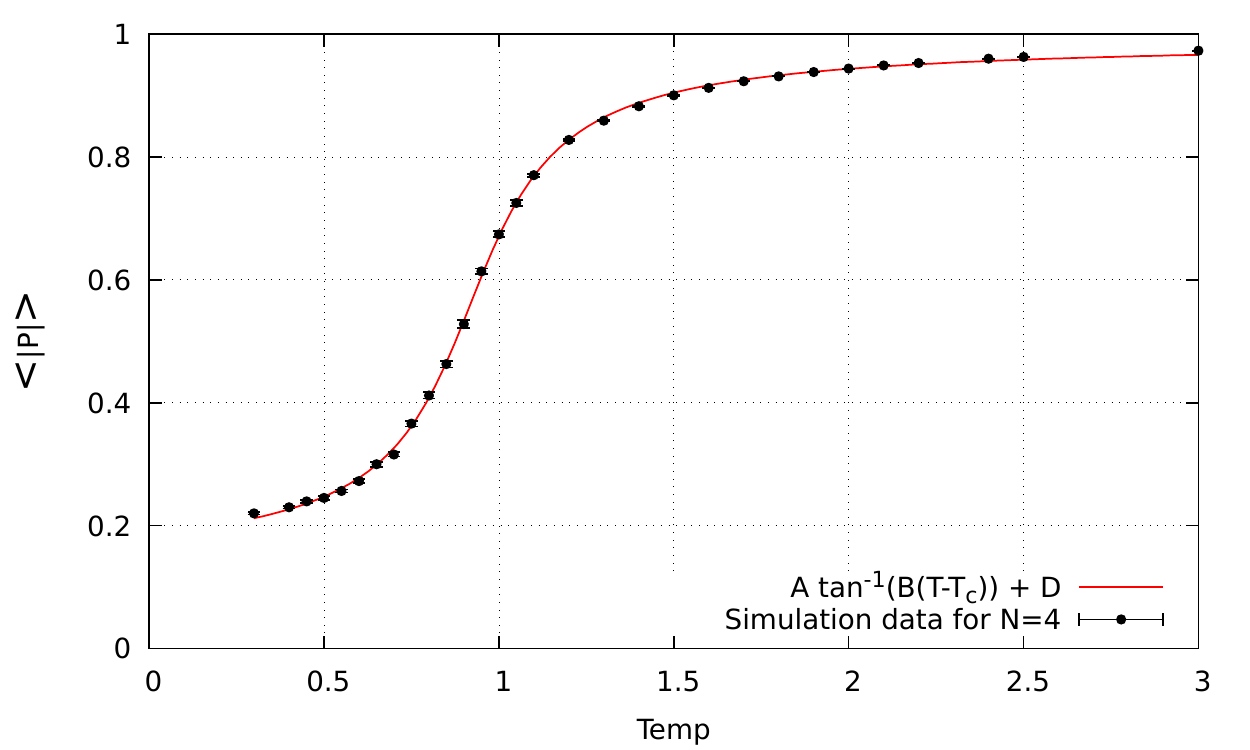}
\end{center}
\caption[Polyakov loop for $\mu = 2.0$ in the bosonic BMN model.]{Plot for the expectation value of the Polyakov loop as a function of the temperature. It indicates the existence of a phase transition around $t \approx 0.91$. Here, $d = 9$, $\lambda = 1$, $N = 4$, $T = 10$ and $\mu = 2$.}
\label{fig:ch8-polyakov}
\end{figure}

We provide the values of the fit parameters $A, B, T_C$ and $D$ in Table. \ref{tb:ch8-polyakov}. A plot for the expectation value of the Polyakov loop as a function of the temperature is given in Fig. \ref{fig:ch8-polyakov}. From the data we see the existence of a phase transition around $t \approx 0.91$. The simulations were performed for $\lambda = 1$, $N = 4$, $T = 10$ and $\mu = 2$.

\begin{table}[hbt!]
\caption[Fit parameters for the Polyakov loop in the bosonic BMN model.]{The values of fit parameters $A,B,T_c$ and $D$. Here, $d = 9$, $\lambda = 1$, $N = 4$, $T = 10$ and $\mu = 2.0$.}
\label{tb:ch8-polyakov}
\begin{center}
\begin{tabular}{ | c | c | } 
\hline
Parameter & Value \\
  \hline \hline
    A	& 0.274143 $\pm$ 0.001743\\
    B	& 5.25498 $\pm$ 0.09958\\
    $T_{c}$	& 0.918857 $\pm$ 0.002204\\
    D   & 0.560791 $\pm$ 0.001572\\
  \hline
\end{tabular}
\end{center}
\end{table}

\subsubsection{Internal energy}

\begin{figure}[hbt!]
\begin{center}
\includegraphics[scale=1.0]{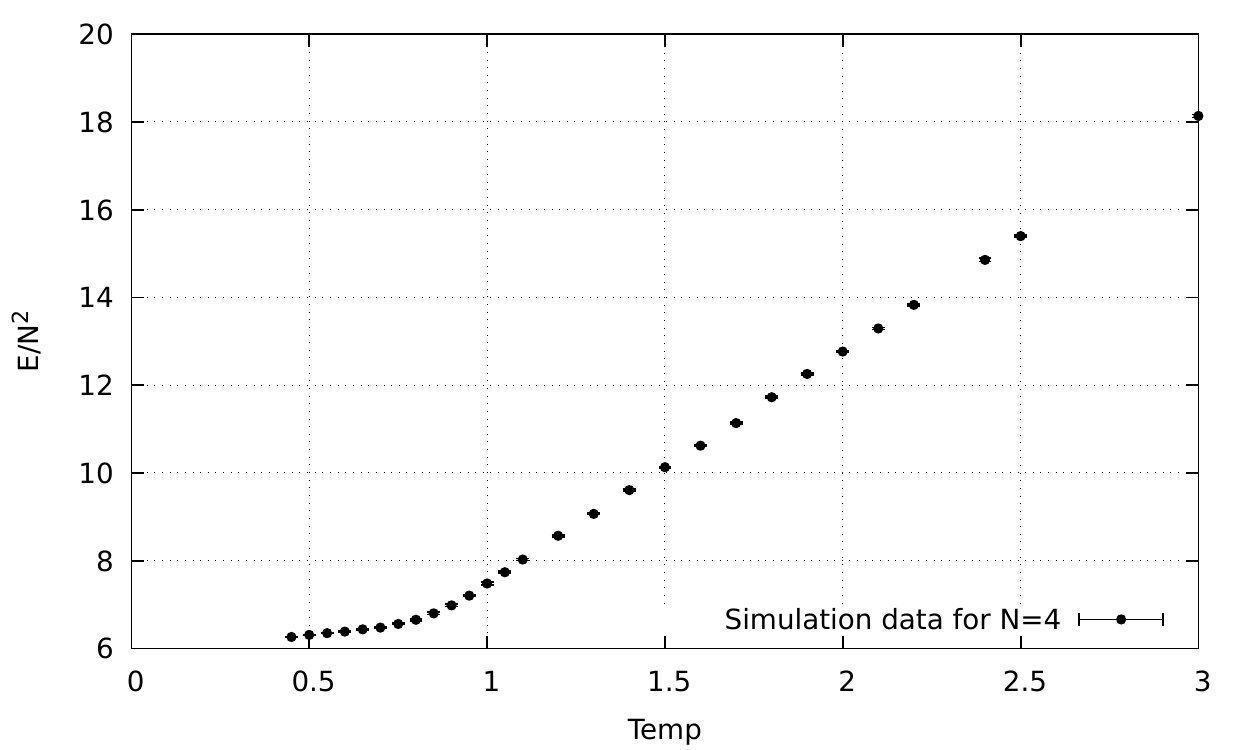}
\end{center}
\caption[Internal energy for $\mu=2.0$ in the BMN model.]{The internal energy is plotted against the temperature for $N=4$ and $\mu=2.0$. Here, $d = 9$, $\lambda = 1$ and $T = 10$.}
\label{fig:ch8-energy}
\end{figure}

In Fig. \ref{fig:ch8-energy} we show the simulation data for the internal energy as a function of the temperature for the $N=4$ case with $\mu = 2.0$, $\lambda = 1$ and $T = 10$.

\subsubsection{Extent of space}

\begin{figure}[hbt!]
\begin{center}
\includegraphics[scale=1.0]{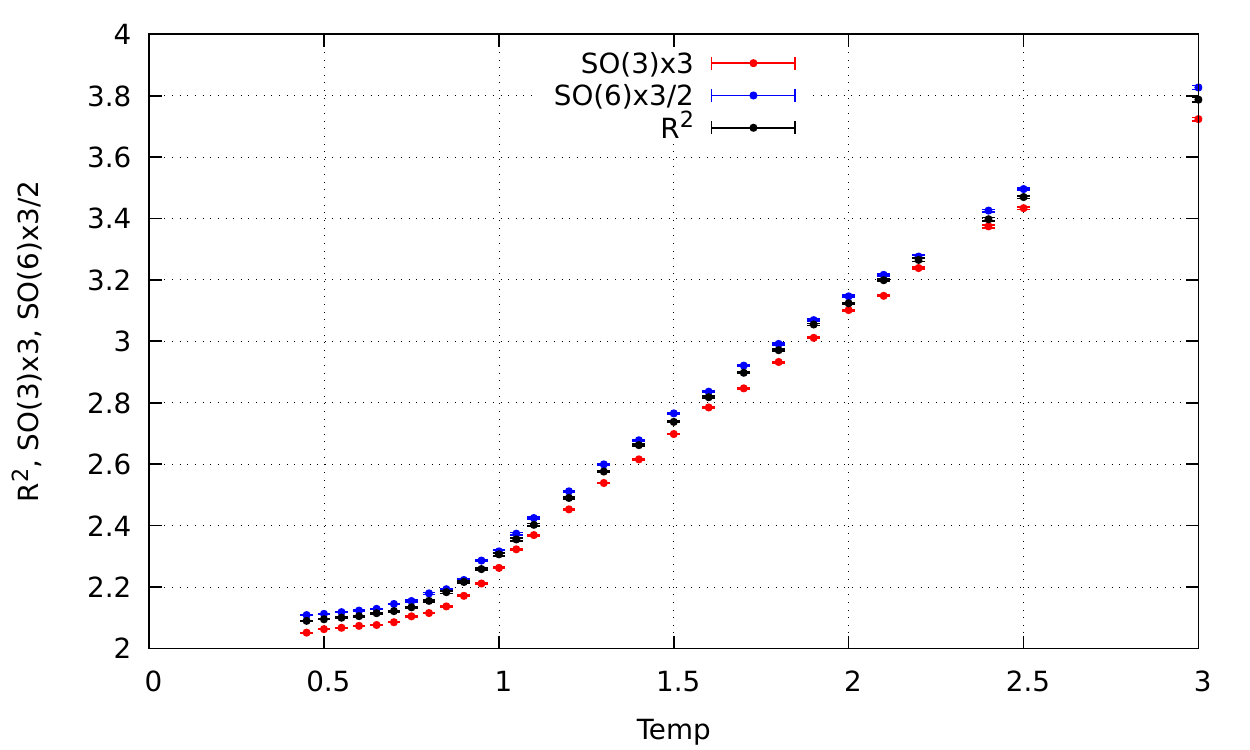}
\end{center}
\caption[Extent of space for $\mu=2.0$ in the bosonic BMN model.]{The extent of space and its components ($SO(3)$ and $SO(6)$) are plotted against the temperature for $N=4$ and $\mu=2.0$. Here, $d = 9$, $\lambda = 1$ and $T = 10$.}
\label{fig:ch8-R2}
\end{figure}

In Fig. \ref{fig:ch8-R2} we provide the simulation data for the extent of space and its components ($SO(3)$ and $SO(6)$) against the temperature for the $N = 4$ case with $\mu = 2.0$, $\lambda = 1$ and $T = 10$.

\subsubsection{Myers term}

\begin{figure}[hbt!]
\begin{center}
\includegraphics[scale=1.0]{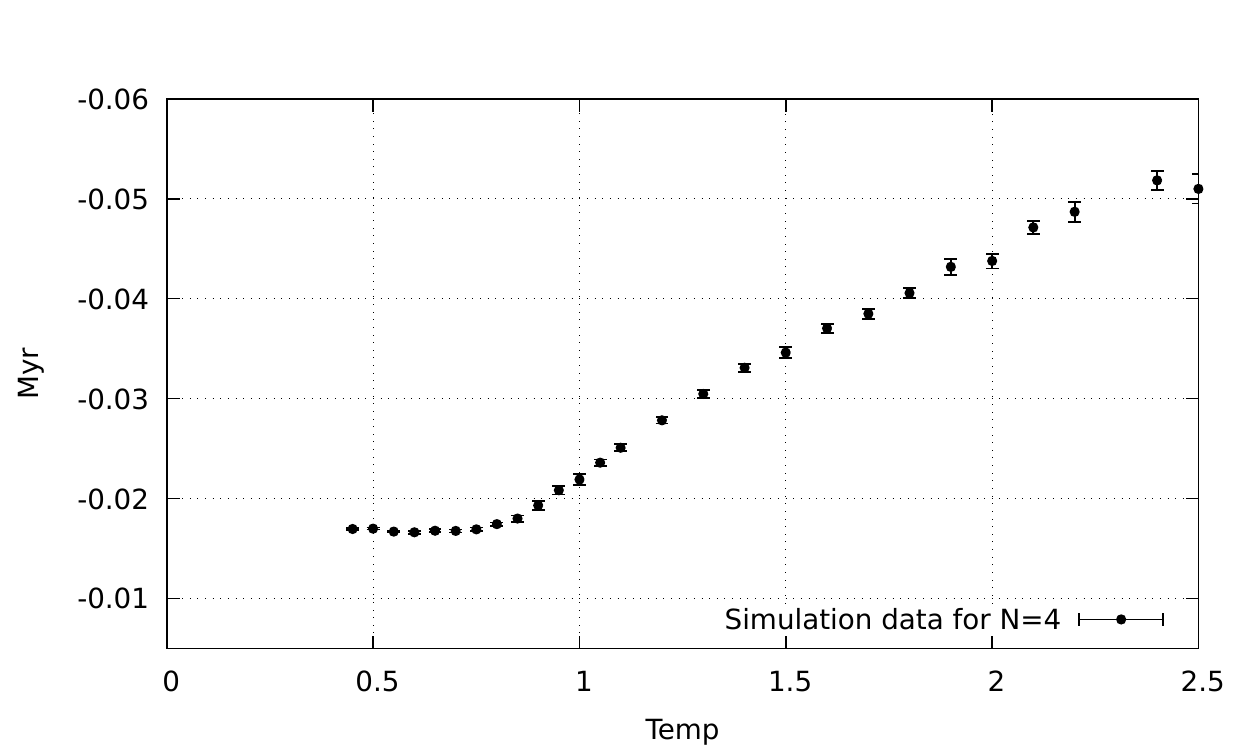}
\end{center}
\caption[Myers term for $\mu=2.0$ in the bosonic BMN model.]{A plot of Myers term against the temperature for $N=4$ and $\mu=2.0$. Here, $d = 9$, $\lambda = 1$ and $T = 10$.}
\label{fig:ch8-myr}
\end{figure}

In Fig. \ref{fig:ch8-myr} we provide the Myers term observable against the temperature for the case $N = 4$ with $\mu = 2.0$, $\lambda = 1$ and $T = 10$.

\subsection{Parametrized phase diagram}

\begin{figure}[hbt!]
\begin{center}
\includegraphics[scale=1.0]{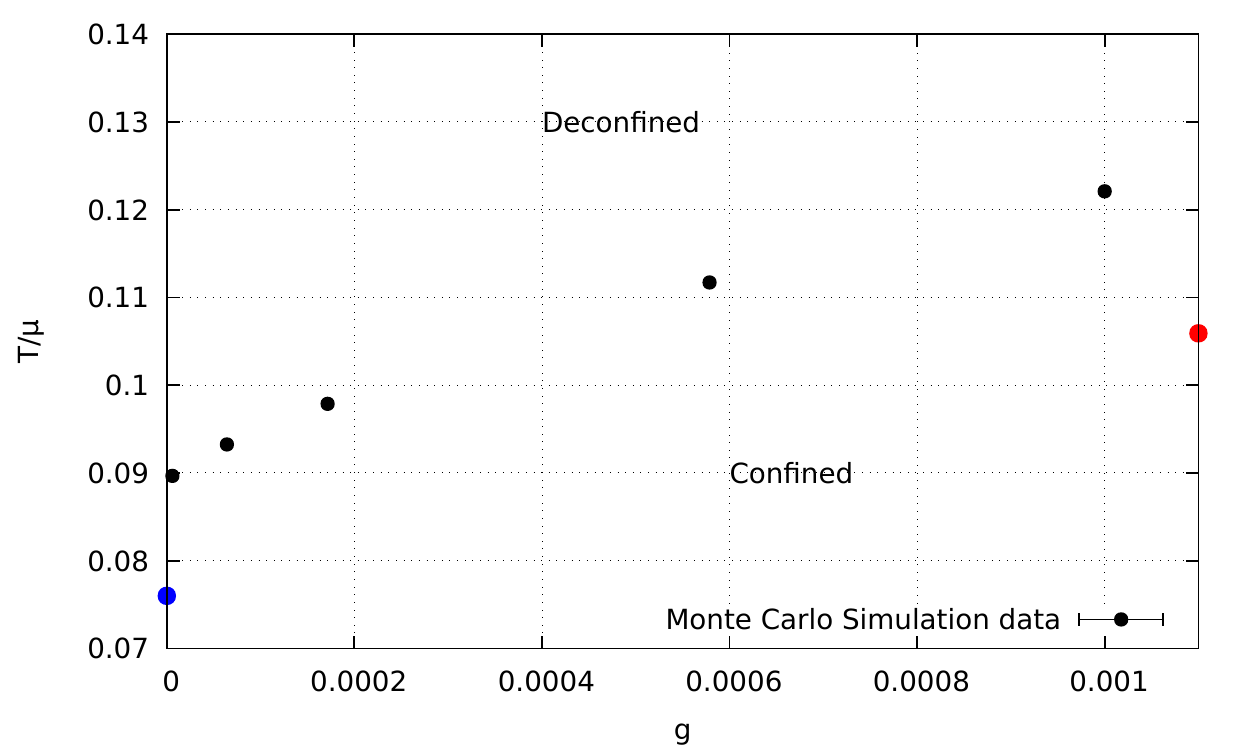}
\end{center}
\caption[Parametrized phase diagram of the bosonic BMN model.]{The phase diagram for the bosonic BMN model. For each value of $g$, the critical temperature, $T_C$ is evaluated by fitting the Polyakov loop data with the functional form given in Eq. \eqref{eq:ch6-fit}. The colored circles represent the values of $T_C/\mu$ in the weak ($g \ll1$) and the strong ($g \gg 1$) coupling limit for the full BMN model. In each case, we set $\lambda = 1$, $N = 4$, $T = 10$ and $d = 9$.}
\label{fig:ch8-phase_str}
\end{figure}

As discussed in the Chapter 1, a two dimensional parametrized phase diagram can be built for the model by using a dimensionless temperature $T/\mu$ and a dimensionless coupling $g = \lambda/{\mu}^{3}$. This phase diagram is generated by evaluating the critical temperature for different values of the dimensionless coupling, $g$. Fig. \ref{fig:ch8-phase_str} shows the phase diagram obtained after simulating the model for five distinct values of $g$. The data used in the plot is given in Table. \ref{tb:ch8-phase-str}. \\

\begin{figure}[hbt!]
\begin{center}
\includegraphics[scale=1.0]{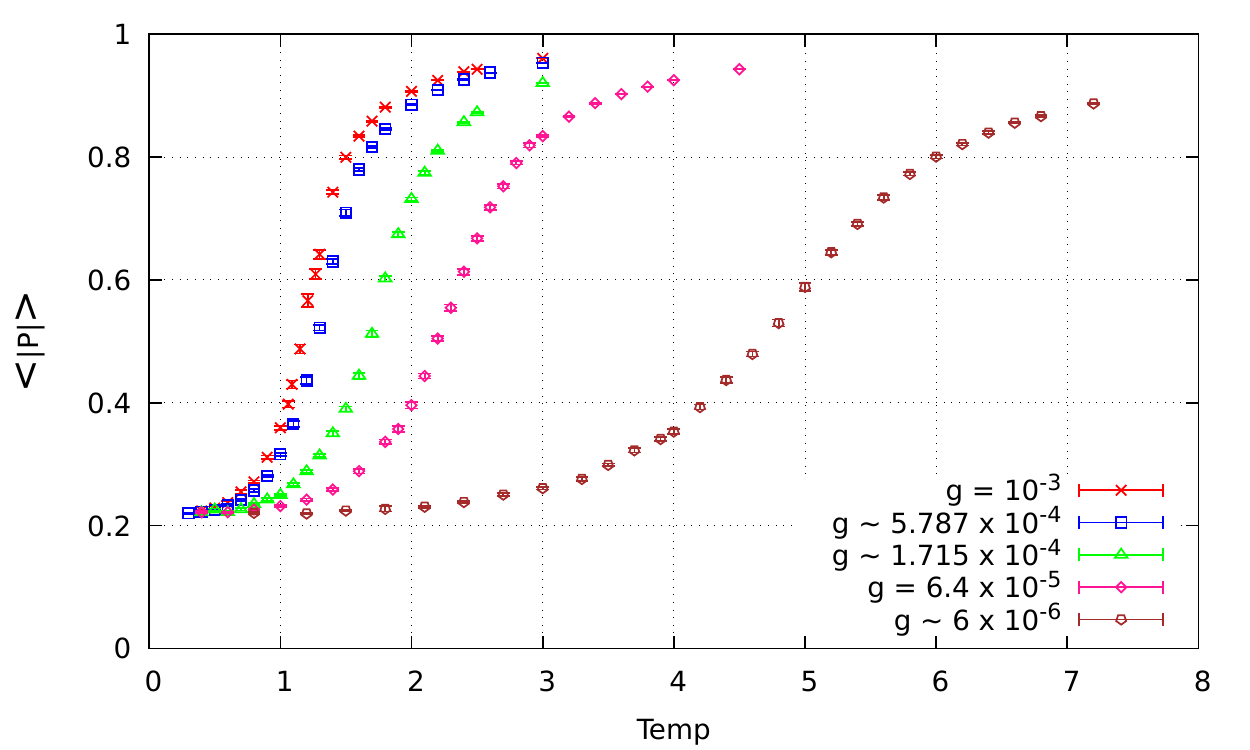}
\end{center}
\caption[The Polyakov loop observable for different $g$ values.]{A representative set of Polyakov loop data that is used to determine the critical $T /\mu$ of the deconfinement transition. We set $\lambda = 1$, $N = 4$, $T = 10$ and $d = 9$.}
\label{fig:ch8-coupling_polyakov}
\end{figure}

In Fig. \ref{fig:ch8-coupling_polyakov} we show the expectation values of the Polyakov loop against temperature for these five $g$ values.

\begin{table}[hbt!]
\caption[Data for the parametrized phase diagram of the bosonic BMN model.]{This table shows the obtained values of $T_C/\mu$ for different values of the coupling $g$. The corresponding $\mu$ values are also provided. For all cases we used $\lambda = 1$, $N = 4$, $T = 10$ and $d = 9$.}
\label{tb:ch8-phase-str}
\begin{center}
\begin{tabular}{ | c | c | c | } 
\hline
$\mu$ & $g=\frac{\lambda}{\mu^{3}}$ & $\frac{T_{c}}{\mu}$ \\
  \hline \hline
10  &   0.001000000	&   0.122101 $\pm$ 0.000238\\
12  &   0.000578703	&	0.111706 $\pm$ 0.0001944\\
18  &   0.000171467	&	0.0978775 $\pm$ 0.0002242\\
25  &   0.000064000	&	0.093240 $\pm$ 0.000153\\
55  &   0.000006010	& 0.0896591 $\pm$ 0.0001183\\
  \hline
\end{tabular}
\end{center}
\end{table}

\chapter{Conclusion}
\noindent The main goal of this thesis was to recover the parameterized phase diagram for the bosonic BMN matrix model using Monte Carlo simulations. \\

\noindent Initially, we cross checked our simulation results with the existing results for the case of a smaller model, the $D = 4$ model, using hybrid Monte Carlo algorithm. Next, we studied the quenched BFSS model using various observables. We found that our results were in good agreement with those of the earlier studies. Further, we simulated the quenched BMN matrix model for the mass parameter $\mu = 2.0$ and obtained the value of critical temperature. Our result matches exactly with the one given in Ref. \cite{Asano:2020yry}. Finally, we performed simulations to obtain the parametrized phase diagram for the bosonic BMN model, which was our main objective. \\

\noindent The primary observable used to investigate the deconfinement phase transition in the model is the Polyakov loop. It can be clearly seen from the simulation data that the system undergoes a deconfinement phase transition. The three other observables, the internal energy, extent of space and Myers term, were computed to examine the value of critical temperature obtained from the Polyakov loop. The final calculations were done for the $N = 4$ case. For large $N$, the Polyakov loop will have the similar behaviour as seen from the $N = 4$ data. The only difference would be that at the critical point the observable would sharply change its value. \\

\noindent An important extension of this work would be to add fermions to the model and see how the results change.

\bibliography{main}
\printindex
\end{document}